\documentclass[%
reprint,
superscriptaddress,
nofootinbib,
aps,
pra,
longbibliography
]{revtex4-1}

\usepackage{graphicx} 
\usepackage{dcolumn} 
\usepackage{bm} 
\usepackage{braket}
\usepackage{amsmath, amsthm}
\usepackage{bm}
\usepackage{amssymb}
\usepackage{setspace}
\usepackage[euler]{textgreek}
\allowdisplaybreaks
\usepackage{bbm}
\usepackage{float}
\usepackage{multirow}
\usepackage{tabularx}
\usepackage{subcaption}
\newcommand{\epsiloneff}{\epsilon_{\text{eff}}}

\newtheorem{proposition}{Proposition}

\DeclareMathOperator*{\argmax}{arg\,max}

\usepackage[colorlinks=true,bookmarks=false,citecolor=blue,urlcolor=blue]{hyperref}

\begin{document}

\title{Quantum Network Utility: A Framework for Benchmarking Quantum Networks}

\author{Yuan Lee}
\email{leeyuan@mit.edu}
\thanks{Equal contribution}
\affiliation{Department of Electrical Engineering and Computer Science, Massachusetts Institute of Technology, Cambridge, Massachusetts 02139, USA}

\author{Wenhan Dai}
\email{whdai@mit.edu}
\thanks{Equal contribution}
\affiliation{Quantum Photonics Laboratory, Massachusetts Institute of Technology, Cambridge, Massachusetts 02139, USA}
\affiliation{College of Information and Computer Sciences, University of Massachusetts, Amherst, MA 01003, USA}

\author{Don Towsley}
\affiliation{College of Information and Computer Sciences, University of Massachusetts, Amherst, MA 01003, USA}

\author{Dirk Englund}
\email{englund@mit.edu}
\affiliation{Department of Electrical Engineering and Computer Science, Massachusetts Institute of Technology, Cambridge, Massachusetts 02139, USA}
\affiliation{Research Laboratory of Electronics, Massachusetts Institute of Technology, Cambridge, Massachusetts 02139, USA}

\date{October 14, 2022}

\begin{abstract}

The absence of a common framework for benchmarking quantum networks is an obstacle to comparing the capabilities of different quantum networks. We propose a general framework for quantifying the performance of a quantum network, which is based on the value created by connecting users through quantum channels. In this framework, we define the quantum network utility metric $U_{QN}$ to capture the social and economic value of quantum networks. While the quantum network utility captures a variety of applications from secure communications to distributed sensing, we study the example of distributed quantum computing in detail. We hope that the adoption of the utility-based framework will serve as a foundation for guiding and assessing the development of new quantum network technologies and designs.

\end{abstract}

\maketitle

\section{Introduction}

Quantum networks transmit quantum information between quantum systems separated by large distances, enabling applications like quantum cryptography and quantum sensing that are not possible with classical communication networks alone~\cite{Wehner_2018,Kimble_2008}. Efforts are underway across the globe to develop the cornerstones of such quantum networks, with the goal of distributing quantum information between quantum memories. These efforts are diverse, employing a range of protocols and hardware to support long-distance, unconditionally secure communication, precision sensing/navigation and distributed quantum computing. On-demand quantum entanglement is now possible between separated quantum memories~\cite{Humphreys_2018} and quantum memories have been shown to offer a clear advantage in the quantum secure information capacity compared to direct transmission over equivalent loss channels~\cite{Bhaskar_2020}. Recent theory established the secret key capacity for arbitrary quantum communication networks~\cite{Pirandola_2017,Pirandola_2019}, and various quantum network routing protocols are being developed~\cite{Muralidharan_2016,Pant_2019,Dahlberg_2019,router_2020} to try and approach these capacities.

But given this multitude of applications, is there a way to quantify the usefulness of a quantum network? 
Even though Ref.~\cite{Pirandola_2019} established the maximum point-to-point quantum communication capacity, it considers channel losses as the only limit on quantum communication rates. In practice, local operation errors and sub-optimal link layer network protocols can also limit the performance of quantum networks. Furthermore, we would like to benchmark quantum networks with respect to other network tasks, such as computing and sensing. In the absence of a commonly agreed-upon framework that can accommodate various quantum network-enabled applications, comparing the capabilities of different quantum networks at various global network tasks remains difficult.

Fundamentally, the value of a network derives from the applications it enables by connecting people and things. A telephone network's worth derives from the utility seen by the callers it connects. The value of a datacenter network derives from the added utility of networked, rather than isolated, computers. A wireless sensor network enabled by 5G technology adds value by connecting sensors that, taken in isolation, would be less valuable: i.e. the utility of the whole is greater than the sum of its parts. 

\begin{figure}
    \centering
    \includegraphics[width = 0.8 \linewidth]{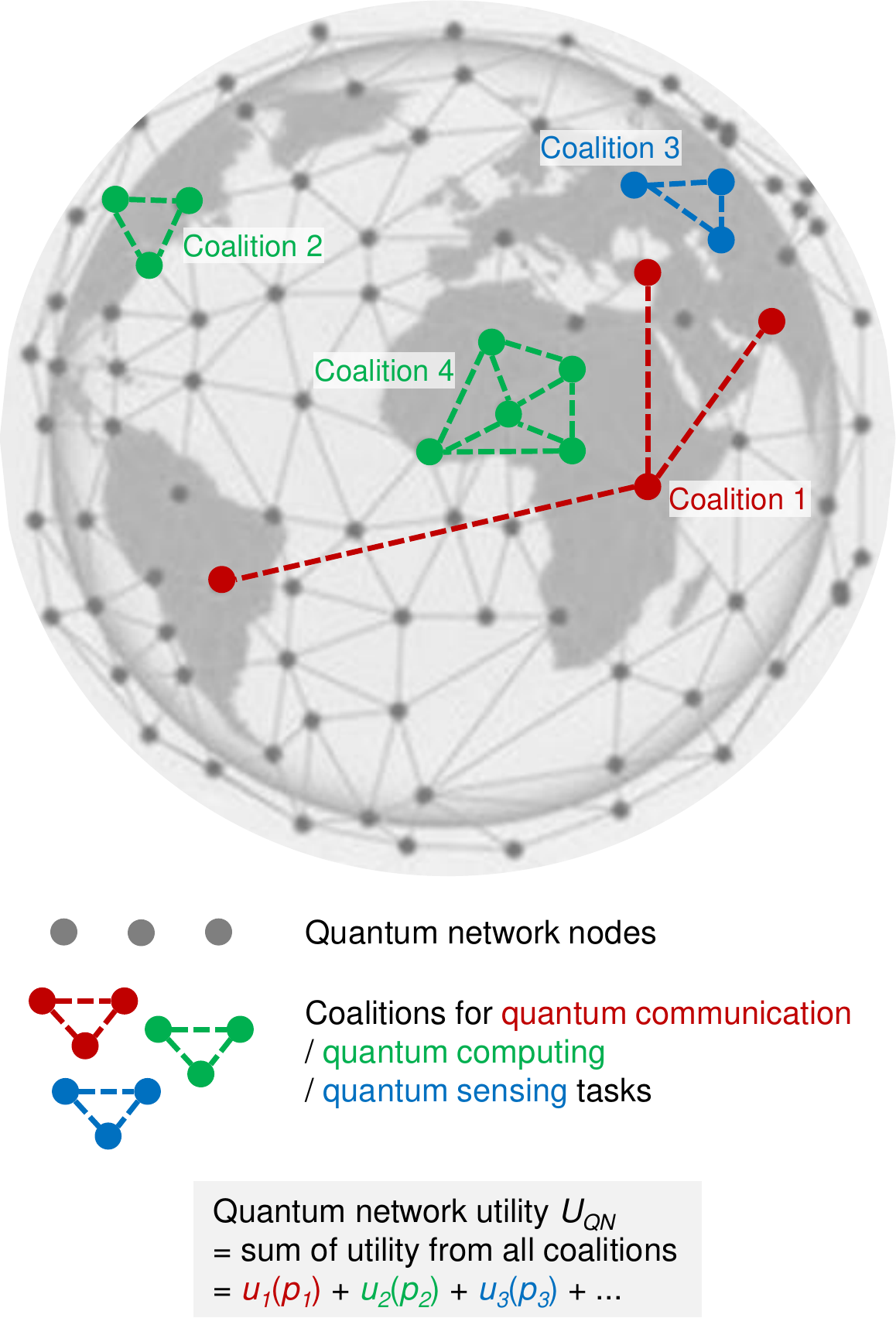}
    \caption{Illustration of a global quantum network. Nodes are connected by quantum channels. Different coalitions (i.e. subsets) of nodes in a quantum network perform different tasks repeatedly. Each task is associated with a utility function. The performance of a quantum network is measured by the aggregate utility provided by the network.}
    \label{fig:illustration}
\end{figure}

When it comes to quantifying the worth of a quantum network, we are guided by the same principle: we consider how the ability to pass quantum information between devices or people creates new value. 
Specifically, we introduce utility-based metrics to quantify the performance of a quantum network in servicing the diversity of envisioned quantum network applications. These metrics form a general framework for comparing the utility of different quantum networks, such as those illustrated in Fig.~\ref{fig:illustration}. These metrics can also be used to guide the design of quantum networks, so they can best serve the quantum information needs of their users.

Analyzing these metrics reveals new insights in the form of scaling laws for the performance of quantum networks. These laws serve the same purpose for quantum networks as Metcalfe's law does for classical networks: they provide network designers and users with a practical measure of network utility, which in turn informs the expansion of existing networks or the construction of new ones. In addition, these scaling laws can also serve the same purpose for quantum networks as Moore's law does for classical computing: they provide a measure of progress for quantum networks in their applications of communication, computing or sensing.

After explaining the general framework for quantifying the performance of a quantum network, we propose one example of a quantum network metric that extends IBM's quantum volume~\cite{Cross_2019} for benchmarking quantum computers to quantum networks.\footnote{We provide some code to compute the example metric in an interactive notebook, which can be found at \url{https://tinyurl.com/quantumNetworkUtility2022}.} This framework can also be used to construct metrics for other common quantum network tasks. By incorporating information on the network's hardware, connectivity and link layer protocols, these metrics provide realistic and physically relevant measures of a quantum network's performance.

Our utility-based model for analyzing quantum networks is similar in spirit to the approach Refs.~\cite{Kelly_97}~and~\cite{Kelly_98} take to analyze classical networks. These papers focus on an efficient framework for rate control, whereas we focus on valuing and comparing networks. Nevertheless, our analysis can potentially be extended to the development of a rate control algorithm.

\section{Benchmarking Quantum Networks}

Quantum networks establish quantum communication channels between distant nodes, who can use these channels to perform quantum computation, distribute secure keys or send quantum states. For concreteness, we assume that the quantum network establishes such channels by distributing entanglement between end users, but our framework also applies to other quantum communication protocols.

A fundamental description of a quantum network's capabilities is given by its rate region, which describes the communication rates it can simultaneously enable among all groups of users. The rate region captures the scarcity of entanglement resources, as an entangled state that is used to connect one group of users cannot also be used to connect another group of users. The rate region incorporates information on the link-layer and network-layer protocols~\cite{Dahlberg_2019, Pant_2019} of the quantum network, such as the routing algorithms between nodes and the error-correction procedures at individual nodes. 

The quantum communication enabled by the network can be applied to a series of tasks from which users of the network derive value. We attribute a utility function to each of the tasks the quantum network performs. It will be convenient to think of this utility as the users' willingness to pay for the output of these tasks, but more generally, this abstract ``utility" can reflect a monetary value, an equivalent cost of classical cryptography, or an abstract social good. Then the performance of a quantum network is summarized by the maximum aggregate utility the network can provide. 

The rate region alone is poorly suited for comparing the performance of quantum networks, because the rate region may not reflect the reality that some communication channels are more valuable than others. Moreover, different quantum networks may connect different groups of users, so their rate regions are incomparable. Perhaps most importantly, quantum networks may be used for different applications, and the performance of a quantum network must be evaluated in the context of the application for which it is used. The aggregate utility metric incorporates information about how value is derived from the quantum network by assigning utilities to tasks that can be completed using quantum communication. Not only can these utilities guide the design of quantum network protocols, but they can also be used to enforce a fair entanglement sharing policy between groups of users.

We introduce some notation for concreteness. Let the quantum network be described by the graph $(\mathcal{V}, \mathcal{E}_p)$, where $\mathcal{V}$ is the set of nodes and $\mathcal{E}_p$ is the set of physical links connecting pairs of nodes. The edge set $\mathcal{E}_p$ also describes the set of multiparty communication channels the quantum network can enable without performing entanglement swaps. This entanglement can be used to perform tasks, each of which involves a subset of nodes in $\mathcal{V}$. (Note that tasks and channels are distinct concepts.)

Suppose there are $D$ tasks the quantum network performs repeatedly. The quantum network completes tasks as frequently as the output entanglement rates allow. Let the feasible task region $\mathcal{W}$ be the set of task completion rate vectors $P = (p_1, p_2, \ldots, p_D)$ that can be sustained by the quantum network. Here, $p_i$ is the rate at which the $i^{\text{th}}$ task is completed. A vector of task completion rates can be sustained if the rate at which the quantum network must consume entanglement lies within the rate region of the network.

Users derive value from the quantum network when tasks are completed. In our proposed framework, we allow the utility derived from a task to depend nonlinearly on the rate at which this task is completed, but the utility users enjoy from one task is independent of the rates at which other tasks are completed.

Let $u_i:\mathbb{R}_+ \rightarrow \mathbb{R}$ be the utility function associated with the $i^{\text{th}}$ task. If the quantum network performs task $i$ at rate $p_i$, users derive utility $u_i(p_i)$ from this task. Thus, if the network achieves the task completion rate $P = (p_i)_{i=1}^D$, users derive an aggregate utility $\sum_{i=1}^D u_i(p_i)$ over all tasks.

We choose the task completion rate vector that maximizes the aggregate utility users derive from the network. Note that the feasible task region implicitly depends on the rate region of the network. The quantum network's performance is measured by the following metric, known as the quantum network utility:
\begin{align}
    U_{QN} & = \text{ maximum aggregate utility that can be} \nonumber \\
    & \quad \quad \text{provided by the network} \nonumber \\
    &= \max_{P \in \mathcal{W}} \sum_{i=1}^D u_i(p_i). \label{eqn:qnu_definition}
\end{align}

Different applications of quantum networks call for different tasks and utility functions, thereby giving different quantum network metrics. If a quantum network has multiple applications, the set of tasks should include relevant tasks across all applications. Regardless, the metric is defined by an optimization problem over the rate region of the network.

The feasible task region $\mathcal{W}$ can also account for a rate-fidelity tradeoff through the rate region. Supplementary Note~1 provides more details on the connection between the rate region and the feasible task region.

Moreover, when the quantum network utility is treated as a dimensioned quantity, it is a universal measure of the value provided by any quantum network. More details can be found in Supplementary Note~2.

\section{Quantum Network Utility for Distributed Quantum Computation} \label{sec:qnu_computing}

In this section, we present a quantum network metric that applies the quantum volume proposed in~\cite{Cross_2019} to the aggregate utility framework described above. This utility metric $U_{\text{comp}}$ quantifies the value derived from performing distributed quantum computing tasks, and can be viewed as a ``quantum volume throughput".

\subsection{Quantum volume}
The quantum volume is defined with respect to a computing task known as Heavy Output Generation (HOG). The HOG task comprises some number of layers $d$. If $m$ memories are involved in the HOG task, each layer performs $\lfloor m/2 \rfloor$ SU(4) operations over pairs of memories, chosen uniformly at random without replacement. A quantum device is said to perform the $m$-memory HOG task up to depth $d$ if the circuit can be implemented with a given minimum overall accuracy. The value associated with performing an $m$-memory, depth-$d$ HOG task is
\begin{equation*}
    v = \beta^{\min(m, d)},
\end{equation*}
where $\beta = 2$ in Ref.~\cite{Cross_2019}.

The quantum volume is then the maximum value over all HOG tasks that can be performed by the device. Some memories that are otherwise available for computation may be excluded from the HOG task in order to reduce the error per layer, thereby allowing a deeper and thus higher-value HOG task to be performed.

Evaluating the quantum volume of a quantum device is complicated in general, but Ref.~\cite{Cross_2019} approximates the quantum volume by stating that an $m$-memory, depth-$d$ HOG task can be performed only if
\begin{equation} \label{eq:vol_approx}
    md \leq \frac{1}{\epsilon_\text{eff}},
\end{equation}
where $\epsilon_\text{eff}$ is the average error probability per two-qubit gate.

The quantum volume can be interpreted as the cost of classical computing needed to simulate the equivalent quantum circuit.

\subsection{Extension to networks} \label{subsec:utilitydef}

Now we extend the quantum volume to a network setting. This quantum network utility metric differs from the quantum volume in two main ways.
\begin{enumerate}
    \item The quantum network utility explicitly considers the rate at which multi-qubit operations can be performed, because transmitting qubits between nodes incurs a significant time cost.
    \item The quantum network utility accounts for the utility derived simultaneously from tasks performed on different parts of the network, not just the task that generates the highest value.
\end{enumerate}

Without loss of generality, we assume that each node in the network has one single-qubit memory allocated for computation. If a physical node in the network has multiple computation memories, we can split this node into multiple virtual nodes connected by appropriate local links, such that each virtual node has one computation memory. As before, let $\mathcal{V}$ be the set of all (possibly virtual) nodes.
We also assume that the network only produces bipartite entanglement. This simplifying assumption happens to be realistic for near-term quantum networks.

In the above aggregate utility framework, a task is defined by a coalition of at least two nodes $\mathcal{M}_i \subseteq \mathcal{V}$ (with $\lvert \mathcal{M}_i \rvert \geq 2$) performing a HOG task of depth $d_{i}$, $i = 1, 2, \ldots, D$.

The utility $u_i(p_i)$ derived from completing a HOG task $(\mathcal{M}_i, d_i) \in \mathcal{D}$ at a given rate $p_i$ is taken to be proportional to the rate of task completion and the quantum volume of the task:
\begin{equation*}
    u_i(p_i) = \beta^{\min(\lvert \mathcal{M}_i \rvert, d_i)} p_i.
\end{equation*}

The network utility is then the maximum aggregate utility that can be achieved over all task completion rates in the feasible task region $\mathcal{W}$:
\begin{equation*}
    U_{\text{comp}} = \max_{P \in \mathcal{W}} \sum_{i=1}^{D} u_i(p_i).
\end{equation*}
Network utility can be interpreted as the quantum volume throughput enabled by the quantum network. Following the interpretation of quantum volume as the equivalent cost of classical computation~\cite{Cross_2019}, we can also think of the above metric as the cost savings enabled by the quantum network.

\subsection{Modelling the feasible task region}

We now provide a simple model for the feasible task region, so that we can calculate the network utility of distributed quantum computation. This model maps the rate region to the feasible task region, then constructs the rate region itself.

To construct the rate region, we follow Ref.~\cite{Dai_2020} and assume that entanglement swapping occurs with a given efficiency $q_c$ in node $c \in \mathcal{V}$. We also assume that, in the absence of any entanglement swaps, the network produces entanglement between $a$ and $b \in \mathcal{V}$ at rate $f_{ab}$. Note that if nodes $a$ and $b$ are not connected by a physical link, then any communication between these nodes must be generated using entanglement swaps, so $f_{ab} = 0$.

Then, the feasible task region can be inserted into the optimization problem for the quantum network utility:
\begin{widetext}
\begin{align} \label{eq:prob}
    U_{\text{comp}} = \max_{(p_i), (r_{ab}), (w_{ab}^{ac})} & \, \sum_{i=1}^{D} \beta^{\min(\lvert \mathcal{M}_i\rvert, d_i)} p_i \\
    \text{subject to } & \, p_i \geq 0 \, \forall \, i, \quad r_{ab} \geq 0 \, \forall \, a, b \in \mathcal{V}, \quad w_{ab}^{ac} \geq 0 \, \forall \, a, b, c \in \mathcal{V}; \nonumber \\
    & \, r_{ab} = \sum_{i=1}^{D} 2 p_i d_i \left\lbrace \begin{aligned}
    \left(\lvert \mathcal{M}_i \rvert - 1\right)^{-1} & \text{if } \lvert\mathcal{M}_i\rvert \text{ is even} \\
    \lvert \mathcal{M}_i \rvert^{-1} & \text{if } \lvert \mathcal{M}_i \rvert \text{ is odd}
    \end{aligned}
    \right\rbrace \mathbbm{1}_{a \in \mathcal{M}_i} \mathbbm{1}_{b \in \mathcal{M}_i} \, \forall \, a, b \in \mathcal{V}; \nonumber \\
    & \, \lvert \mathcal{M}_i \rvert d_i \leq \frac{1}{\epsilon_{\text{eff}}} \, \forall \, i = 1, \ldots, D \text{ such that } p_i > 0; \nonumber \\
    & \, r_{ab} \leq f_{ab} + \sum_{c \in \mathcal{V} \setminus \lbrace a, b \rbrace} q_c \left( \frac{w_{ab}^{ac} + w_{ab}^{bc}}{2} \right) - \sum_{c \in \mathcal{V} \setminus \lbrace a, b \rbrace} \left( w_{ac}^{ab} + w_{bc}^{ab} \right) \, \forall \, a, b \in \mathcal{V}; \nonumber \\
    & \, w_{ab}^{ac} = w_{ab}^{bc} \, \forall \, a, b, c \in \mathcal{V}. \nonumber
\end{align}
\end{widetext}
This optimization problem is a linear program, and the number of variables is polynomial in $\lvert \mathcal{V} \rvert$ and $D$.

Supplementary Note~3 discusses this optimization problem in greater detail.

\subsection{Case study: repeater chains} \label{sec:case_study}

There is a significant remaining issue: the number of possible coalitions, and thus $D$, grows exponentially with the number of nodes $\lvert \mathcal{V} \rvert$ in the network. 
To make the problem tractable, it is desirable to reduce the set of candidate coalitions that needs to be searched to obtain the utility-maximizing solution set of candidate coalitions.

\begin{figure}[t]
\center	
\includegraphics[width=\linewidth, draft=false]{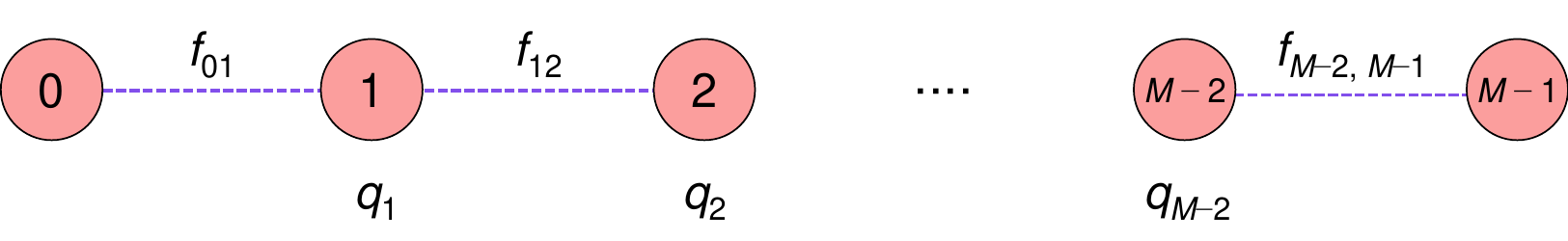}
\caption{Illustration of an $M$-node repeater chain. Physical links are shown as edges between nodes. In this figure, $\mathcal{V} = \{0,1,2,\dotsc, M-1\}$ and $f_{i,j}=0.6 \mathbbm{1}_{\lvert i-j\rvert = 1}$. Relevant coalitions of nodes are $\mathcal{M}_i \subseteq \mathcal{V}$ such that $\lvert \mathcal{M}_i \rvert \geq 2$.}
\label{fig:chain}
\end{figure}

We use repeater chains (Fig.~\ref{fig:chain}) as one example of how we can reduce the set of candidate coalitions when calculating the quantum network utility. In such networks, the connectivity of physical links corresponds to a chain, so that $f_{ab} = 0$ for all non-adjacent $a, b \in \mathcal{V}$.

We say that a coalition is connected if there is a path in the coalition between any two nodes of the coalition. For repeater chains, the following proposition holds.

\begin{proposition}\label{prop:connected_coalition}
In the problem~\eqref{eq:prob}, there exists an optimal solution such that any coalition $\mathcal{M}_i$ with $p_i > 0$ is connected.
\end{proposition}

For a repeater chain with $M = \lvert\mathcal{V}\rvert$ nodes, the number of connected coalitions is $M(M-1)/2$. This is substantially smaller than the number of coalitions in $\mathcal{V}$, which grows exponentially with $M$.

Proposition~\ref{prop:connected_coalition} provides a lower bound on the size of the largest coalition. We now consider homogeneous repeater chains, which are repeater chains with $f_{ab} = f$ for all adjacent nodes $a, b \in \mathcal{V}$ and $q_c = q$ for all nodes $c \in \mathcal{V}$, where $f, q > 0$ are constants.

\begin{proposition}\label{prop:size_largest_coalition}
In a homogeneous repeater chain with perfect quantum memories and no gate errors (i.e. $\epsiloneff = 0$), the size of the largest coalition with nonzero task rate in an optimal solution is bounded from below by
\begin{align*}
M + \log_\beta \frac{M^{\log q} }{(1+q)M^3(M-1)^2/4}.
\end{align*}
\end{proposition}

Proposition~\ref{prop:size_largest_coalition} states that the size of the largest coalition increases as $M - O(\log M)$. Therefore, for large networks with perfect memories and gates, almost all the nodes in the network should be involved in the same coalition, performing a computation task.

We next consider the case where the quantum network produces entanglement of imperfect fidelity. In this case, the size of the largest coalition depends on the fidelity and does not increase as $M-O(\log{M})$. 

\begin{proposition}\label{prop:lb_size_lc_error}
In a quantum network with errors (i.e. $\epsiloneff > 0$), the size of the largest coalition with nonzero task rate in an optimal solution is bounded from above by $\lfloor 1/\sqrt{\epsiloneff}\rfloor$.
\end{proposition}

Proposition~\ref{prop:lb_size_lc_error} states that the size of the largest coalition is dependent on the error rate $\epsiloneff$ and not on the size of the network $M$, assuming the network is sufficiently large. Intuitively, coalitions with size $\lvert \mathcal{M}_i \rvert > 1/\sqrt{\epsilon_{\text{eff}}}$ can only perform HOG tasks up to depth $d_i \leq 1/\sqrt{\epsilon_{\text{eff}}} \lvert \mathcal{M}_i \rvert < \lvert \mathcal{M}_i \rvert$. Completing one such task provides the same utility as completing another task with coalition size and depth $\lvert \mathcal{M}_j \rvert = d_j = d_i < 1/\sqrt{\epsilon_{\text{eff}}}$, but consumes more entanglement. Therefore, it is never optimal to perform HOG tasks over coalitions with more than $1/\sqrt{\epsilon_{\text{eff}}}$ nodes.

\begin{proposition}\label{prop:ub_size_lc_error}
In a homogeneous repeater chain with errors (i.e. $\epsiloneff > 0$), the size of the largest coalition with nonzero task rate in an optimal solution is bounded from below by
\begin{align*}
m+\log_\beta{\frac{4m^{\log_2{q}}\lfloor M/m\rfloor}{(1+q)m^3(m-1)(2M-m+1)}}
\end{align*}
where $m=\lfloor\sqrt{1/\epsiloneff}\rfloor$.
\end{proposition}

Proposition~\ref{prop:ub_size_lc_error} is the equivalent of Proposition~\ref{prop:size_largest_coalition} for repeater chains with errors. It states that for sufficiently large $M$ and sufficiently small $\epsiloneff$, the size of the largest coalition increases as $\lfloor 1/\sqrt{\epsiloneff} \rfloor + O(\log \epsiloneff)$. In other words, the size of the largest coalition stays close to the upper bound of Proposition~\ref{prop:lb_size_lc_error}.

Proofs of these propositions are provided in Supplementary Note~4.

\section{Numerical Results}

\begin{figure}[t]
\center	
\includegraphics[width=\linewidth, draft=false]{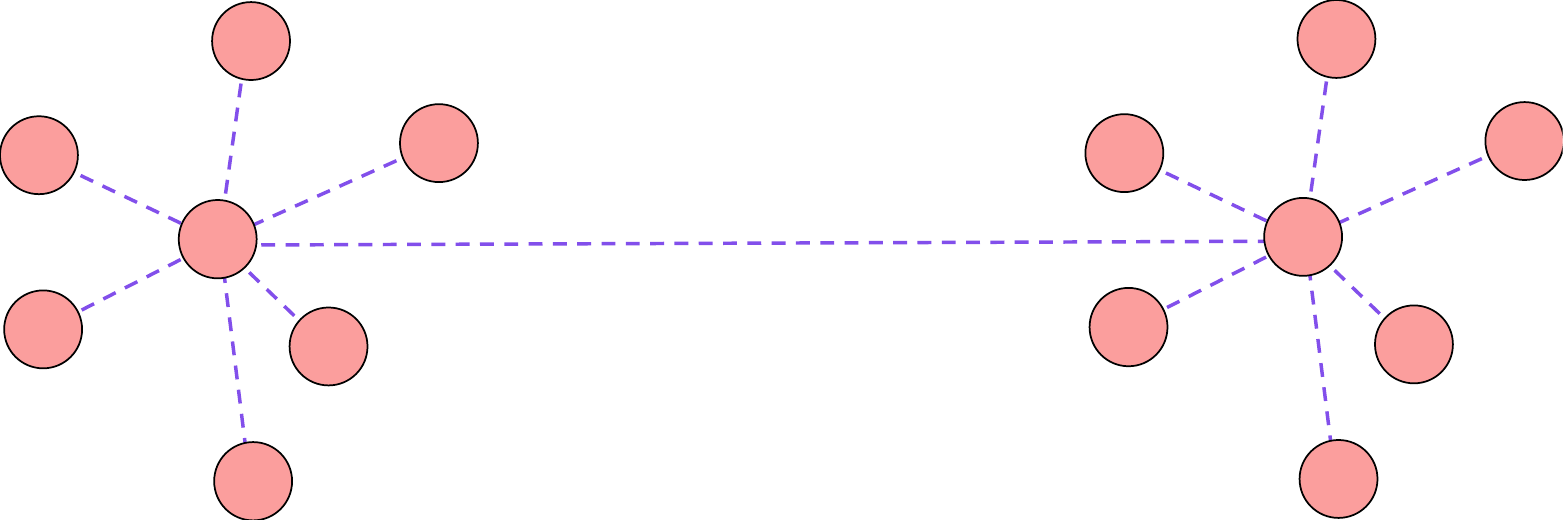}
\caption{Example of a dumbbell network. The long link connecting the two star-shaped subnetworks is the bar; the other links are the spokes.}
\label{fig:dumbbell}
\end{figure}

In this section, we provide numerical results to demonstrate how we can benchmark quantum networks using the quantum network utility. We consider two prototypical quantum networks: homogeneous repeater chains and dumbbell networks. We discussed homogeneous repeater chains in the previous section. A dumbbell network (Fig.~\ref{fig:dumbbell}) comprises two star-shaped networks, whose centers are connected by a physical link. The physical link connecting the centers is known as the bar; the links connecting the star-shaped networks are known as spokes.

We allow the base $\beta$ of each term in the quantum volume to vary to illustrate the effects of varying the utility function. Choosing a larger $\beta$ effectively places a higher value on larger coalitions.

\subsection{Homogeneous repeater chains}

We analyze the homogeneous repeater chain for two values of $\epsilon_{\text{eff}}$: when $\epsilon_{\text{eff}} = 0$, corresponding to the theoretical analysis above, and when $\epsilon_{\text{eff}} = 0.1$, corresponding to a realistic error rate. 
We normalize the rate at which physical links produce entanglement to $f_{i, i+1} = 0.6$, in arbitrary units. We take the efficiency of entanglement swapping to be $q = 0.9$.

\begin{figure}[t]
\center	
\includegraphics[width=0.8\linewidth, draft=false]{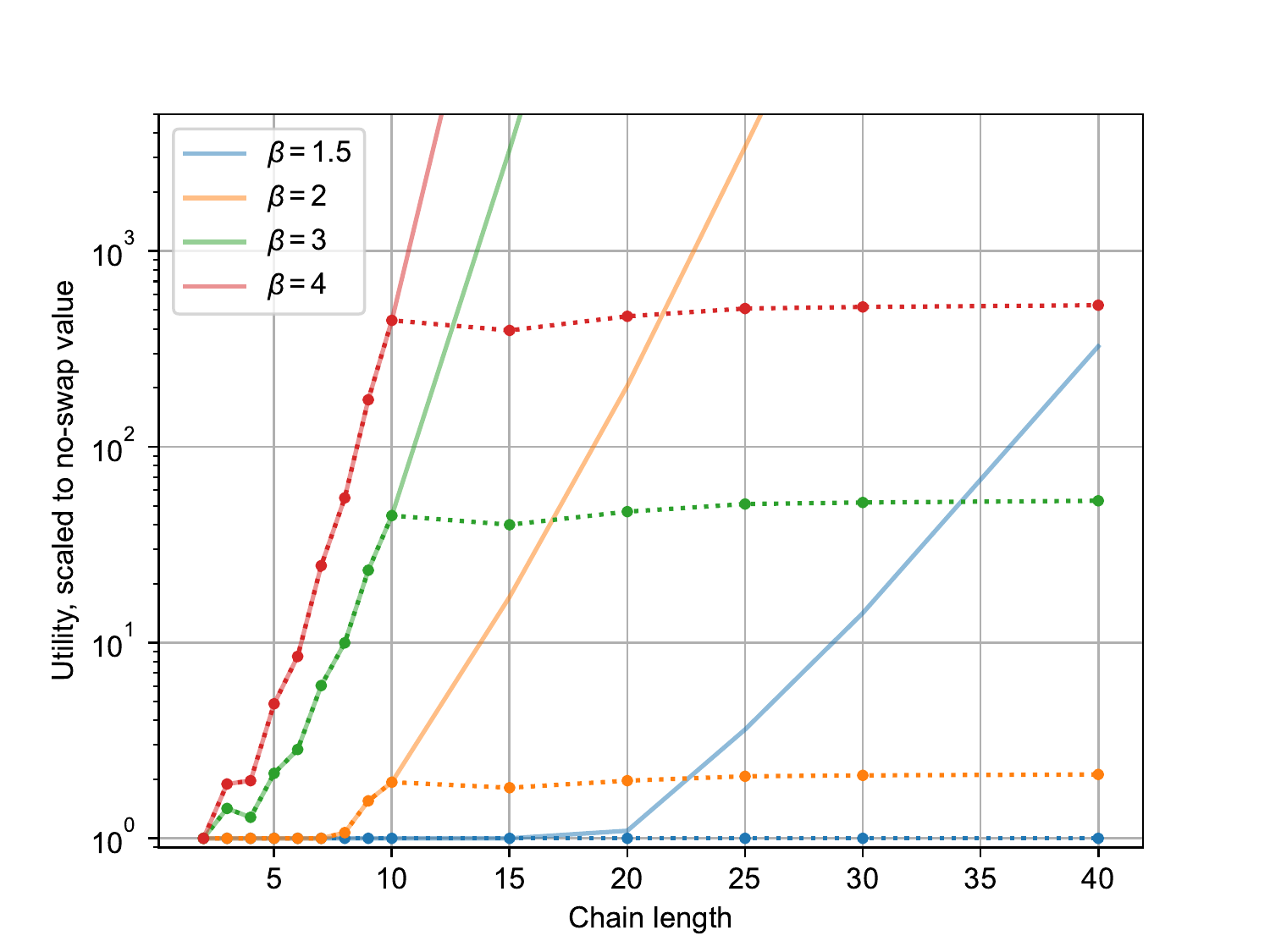}
\caption{Network utility (due to distributed quantum computation) of a repeater chain as a function of chain length $M$, for different bases $\beta$. The network utility is scaled to the aggregate utility when no entanglement swaps are performed (i.e. when entanglement from the physical links is used directly). The solid line corresponds to $\epsilon_{\text{eff}} = 0$; the dotted line corresponds to $\epsilon_{\text{eff}} = 0.01$.}
\label{fig:chain_volume}
\end{figure}

\begin{figure}[t]
\center	
\includegraphics[width=0.8\linewidth, draft=false]{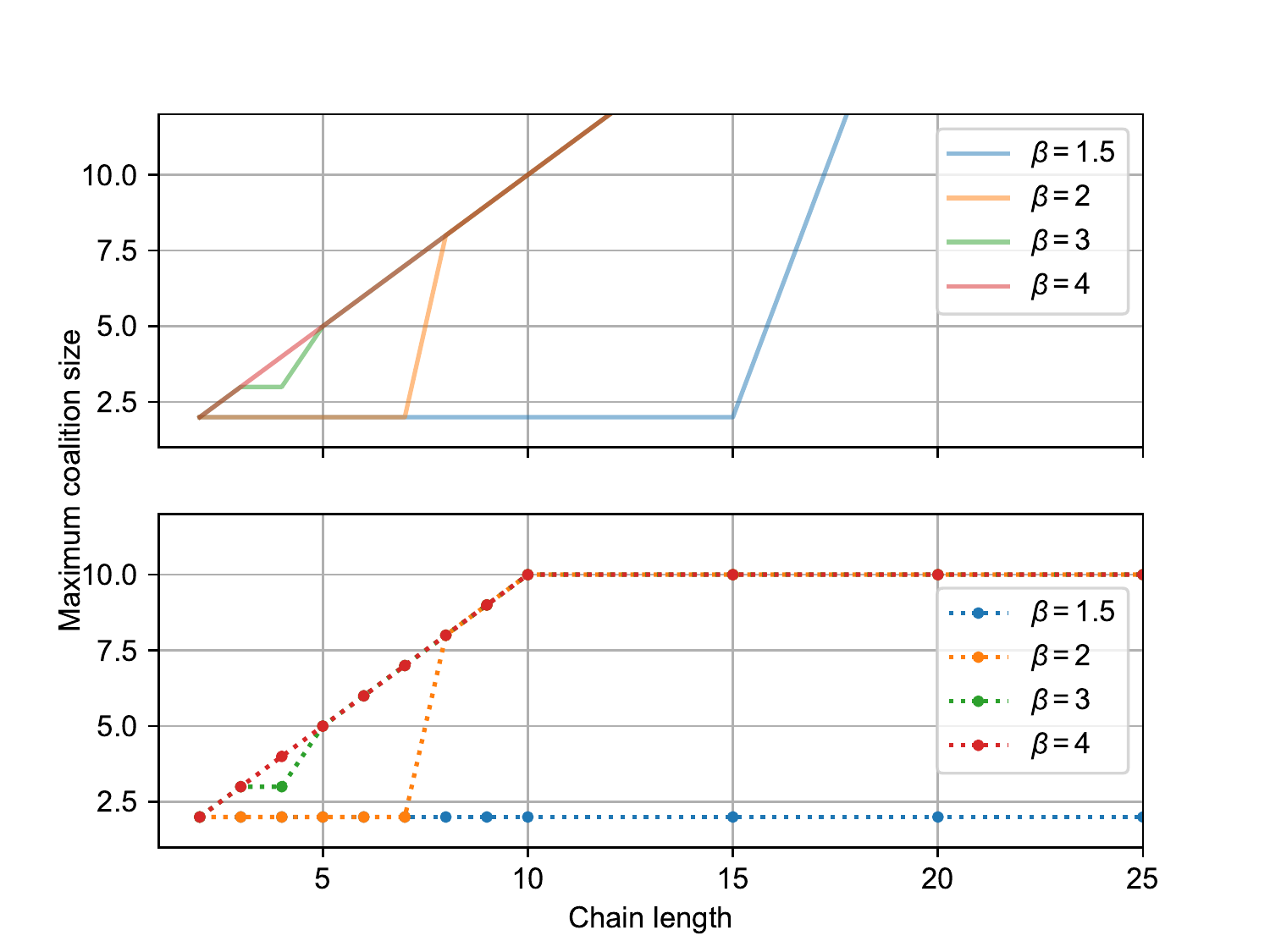}
\caption{Maximum coalition size in a repeater chain as a function of chain length $M$, for different bases $\beta$. The top plot corresponds to $\epsilon_{\text{eff}} = 0$; the bottom plot corresponds to $\epsilon_{\text{eff}} = 0.01$.}
\label{fig:chain_maxcoalsize}
\end{figure}

Fig.~\ref{fig:chain_volume} shows the quantum network utility, as defined in the optimization problem~\eqref{eq:prob}. Fig.~\ref{fig:chain_maxcoalsize} shows the maximum coalition size in the optimal solution to~\eqref{eq:prob}. We observe that with perfect gates ($\epsilon_{\text{eff}} = 0$) and sufficiently many nodes $M$, the largest coalition includes all nodes in the network. However, with imperfect gates ($\epsilon_{\text{eff}} = 0.01$), the optimal solution does not include coalitions with more than 10 nodes. 

\begin{figure}[ht!]
\centering
\begin{subfigure}{\linewidth}
\centering
\includegraphics[width=\textwidth, draft=false]{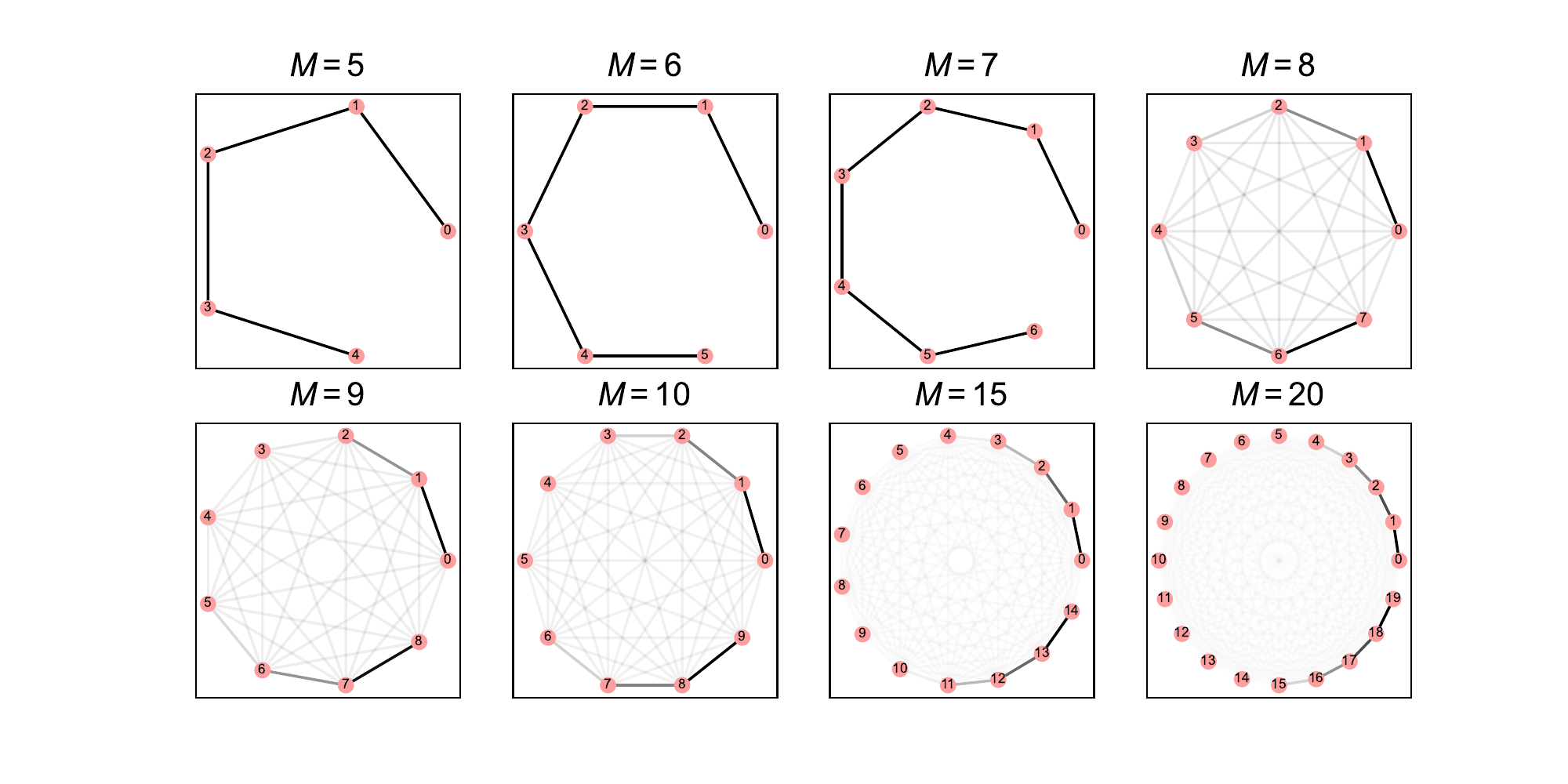}
\caption{$\epsilon_{\text{eff}} = 0$.}
\label{fig:chain_eps0_ratevector}
\end{subfigure}
\begin{subfigure}{\linewidth}
\centering
\includegraphics[width=\textwidth, draft=false]{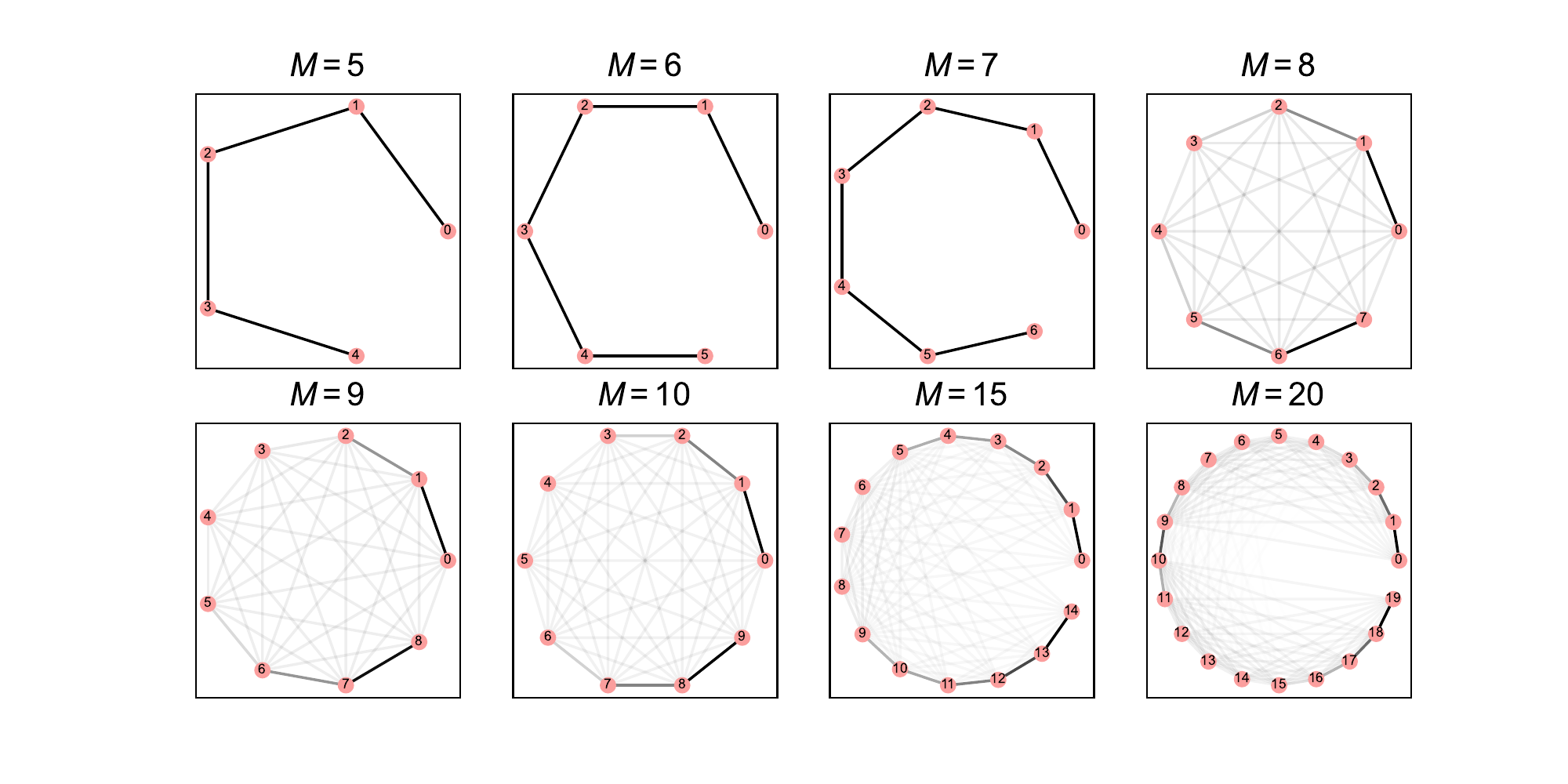}
\caption{$\epsilon_{\text{eff}} = 0.01$.}
\label{fig:chain_eps0.01_ratevector}
\end{subfigure}
\caption{Entanglement graph illustrating the rate vector used to perform HOG tasks, for $\epsilon_{\text{eff}} \in \lbrace 0, 0.01 \rbrace$, $\beta = 2$ and different chain lengths $M$. A darker edge between two nodes means that entanglement between those two nodes is consumed at a higher rate. The nodes of the chain, labeled from $0$ to $M-1$, are arranged in a circle so that the edges of the entanglement graph are shown clearly.}
\end{figure}

In general, by Proposition~\ref{prop:lb_size_lc_error}, the maximum coalition size is bounded from above by $1/\sqrt{\epsilon_{\text{eff}}}$. 
This explains the trends in quantum network utility shown in Fig.~\ref{fig:chain_volume}. When $\epsilon_{\text{eff}} = 0$, the largest coalition spans the full chain, so the quantum network utility increases exponentially with the length of the chain. However, when $\epsilon_{\text{eff}} > 0$, the largest coalition is bounded above by a constant, so in the limit of long chains, the quantum network utility increases linearly with the length of the chain, like the no-swap value. The scaling behavior of the quantum network volume is similar for different choices of the base $\beta$, pointing to the robustness of the quantum network utility for evaluating the performance of a network.

The entanglement graphs in Figs.~\ref{fig:chain_eps0_ratevector} and \ref{fig:chain_eps0.01_ratevector} show the rates at which entanglement between pairs of nodes is consumed in the optimal solution, for repeater chains with $\epsilon_{\text{eff}} = 0$ and $\epsilon_{\text{eff}} = 0.01$ respectively. In the former, we observe that if the chain length is sufficiently large, the entanglement graph is fully connected. As entanglement between nodes 0 and $M-1$ is consumed for a computing task, Proposition~\ref{prop:connected_coalition} implies that there must be a coalition involving all the nodes in the repeater chain. Such behavior is consistent with Proposition~\ref{prop:size_largest_coalition}. Conversely, if the chain length is not sufficiently large, only entanglement between adjacent nodes is consumed.

In the latter, we observe that only entanglement between adjacent nodes is consumed when the chain is sufficiently short, similarly to Fig.~\ref{fig:chain_eps0_ratevector}. As the chain length increases, larger coalitions are formed. However, unlike Fig.~\ref{fig:chain_eps0_ratevector}, the largest coalition is limited to 10 nodes. In particular, when $8 \leq M \leq 10$, the entanglement graph is fully connected, but for $M > 10$, the repeater chain is divided into coalitions of 10 nodes. Hence, Proposition~\ref{prop:size_largest_coalition} can be extended by accounting for how gate errors limit the maximum coalition size. 

\subsection{Dumbbell networks}

We now consider dumbbell networks, and focus on the strength of the bar link relative to the spoke links. A dumbbell network connecting distant subnetworks would have a weaker bar link, which reduces the rate at which full-network computation tasks could be performed. We fix the rate at which physical spoke links produce entanglement to be $f = 0.6$ in arbitrary units, and we vary the ratio of the bar rate to the spoke rate. The dumbbell network is assumed to be balanced, with $M_{\text{side}}$ spoke nodes on each side of the bar. The total number of nodes in the network is $M = 2 M_{\text{side}} + 2$.

\begin{figure}[t!]
\center	
\includegraphics[width=0.8\linewidth, draft=false]{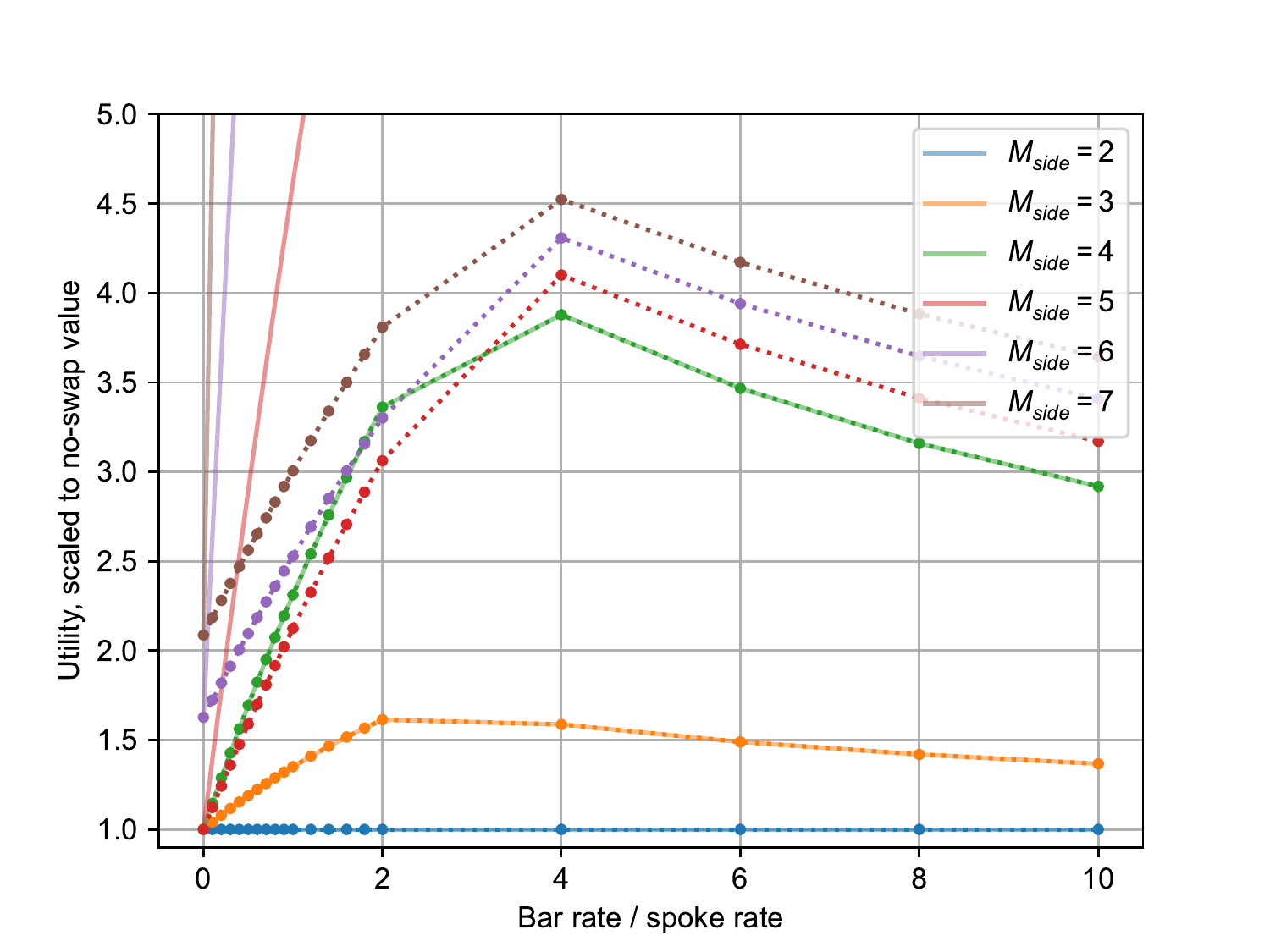}
\caption{Network utility (due to distributed quantum computation) of the dumbbell network as a function of the bar-to-spoke rate ratio. The number of spokes $M_{\text{side}}$ on each side is labeled. The base $\beta = 2$ is used. Each of the spokes generates entanglement at rate 0.6 in the absence of swaps. The volume is scaled to the aggregate utility when no entanglement swaps are performed. The solid line corresponds to $\epsilon_{\text{eff}} = 0$; the dotted line corresponds to $\epsilon_{\text{eff}} = 0.01$.}
\label{fig:dumbbell_volume}
\end{figure}

\begin{figure}[ht!]
\center	
\includegraphics[width=\linewidth, draft=false]{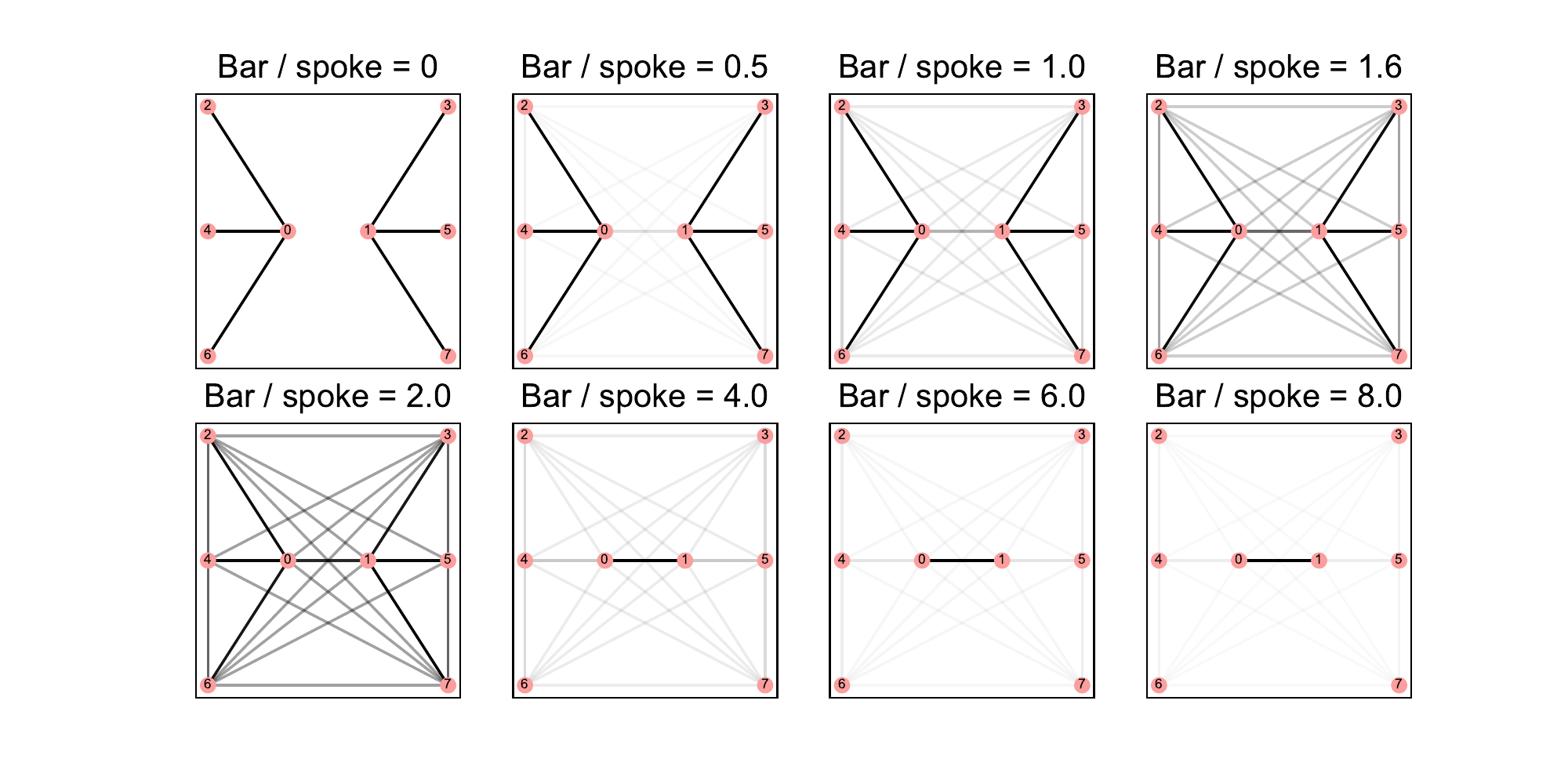}
\caption{Entanglement graph of the dumbbell network illustrating the rate vector used to perform HOG tasks, for $M_\text{side} = 3$, $\epsilon_{\text{eff}} = 0.01$, $\beta = 2$ and different bar-spoke rate ratios. A darker edge between two nodes means that entanglement between those two nodes is consumed at a higher rate. Colors are normalized separately within each entanglement graph. Note that the entanglement graphs for $\epsilon_{\text{eff}} = 0$ are identical to the graphs shown above.}
\label{fig:dumbbell_eps0.01_ratevector}
\end{figure}

Fig.~\ref{fig:dumbbell_volume} shows the quantum network utility of the dumbbell network as a function of the bar rate. As the bar rate increases, we find that network utility initially increases faster than the aggregate utility if no swaps were performed, but the network utility eventually increases more slowly than the no-swap utility. Initially, as the bar rate increases, the network can perform more computation tasks involving the full network, but when entanglement from the spokes is exhausted, additional entanglement along the bar link can only contribute to two-node computations involving the bar nodes. This behavior is reflected in the entanglement graphs in Fig.~\ref{fig:dumbbell_eps0.01_ratevector}. In the limit where the bar rate dominates the spoke rate, the ratio of the network utility to the no-swap utility converges to 1.

We also find that, for sufficiently small networks, the quantum network utility under imperfect gates ($\epsilon_{\text{eff}} = 0.01$) is the same as that under perfect gates ($\epsilon_{\text{eff}} = 0$). In such networks, the coalition that contains all network nodes does not exceed the upper bound on coalition sizes $\lvert M_i \rvert \leq 1 / \sqrt{\epsilon_{\text{eff}}}$. However, for large $M_{\text{size}}$, having imperfect gates restricts the size of the largest coalition that can be formed, thus significantly reducing the quantum network utility. Such behavior mirrors that of the repeater chain (Fig.~\ref{fig:chain_volume}).

\section{Discussion}

The framework we have presented above can be used to answer questions about resource allocation and the commercial viability of quantum networks. It also raises many new questions, like how the utility derived by users of quantum network services can actually be measured, and what market structures could emerge in the quantum network sector as it matures. We briefly address these questions below.

\subsection{Optimal resource allocation}

It is clear that, to maximize the value derived from a quantum network, we should allocate communication resources so as to achieve the optimal task completion rate vector $P^* = (p_i^*)_{i=1}^D = \argmax_{P \in \mathcal{W}} \sum_{i=1}^D u_i(p_i)$. The value derived from the quantum network is then the quantum network utility $U_{QN} = \sum_{i=1}^D u_i(p_i^*)$. 
One major reason to build a quantum network is to exploit the quantum communication it enables. To decide if we should build a quantum network for this reason, we compare the quantum network utility $U_{QN}$ to the total cost of building and operating the quantum network. If $U_{QN}$ exceeds the cost of the quantum network, we should allocate resources to the construction of the quantum network. 

Of course, this cost-benefit analysis ignores other motivations like research and development, which can be significant before quantum networks reach a high technology readiness level. We also assume that the variable costs associated with operating a quantum network are negligible, so that the cost of achieving the task completion rate vector $P^*$ does not depend on the task completion rates themselves.

We are also interested in the allocation of resources between users of the quantum network. A common interpretation of the aggregate utility function $\sum_{i=1}^D u_i(p_i)$ is the willingness to pay of the users who value task completion the most. The optimal allocation of resources between users is then achieved by the following procedure: complete the $i^{\text{th}}$ task for users who enjoy utility exceeding $u_i(p_i^*)/p_i^*$ per unit task completion rate.

\subsection{Commercial viability of a quantum network}

It would be reasonable to leave the provision of quantum network services to the free market when quantum network technologies approach maturity and we seek to exploit quantum communication for commercial applications. We propose a simple model for the private provision of quantum network services. As before, we assume that the operation of a quantum network incurs negligible variable costs.

We consider a market with a monopolistic network operator. This operator is the sole seller of quantum network services. We assume the network operator has no ability to price discriminate.

One option is for the network operator to sell services in the form of elementary entanglement. Users then use this entanglement to perform the tasks from which they derive value. In the simplest pricing scheme, the network operator charges a monthly price of $x$ per unit communication rate. Then a user will purchase entanglement from the network operator if the utility they derive from a task exceeds $x$ times the total amount of entanglement needed for the completion of the task. The network operator will choose to supply entanglement at the price $x$ which maximizes its revenue, and thus profit.

Another option is for the network operator to sell task completions to users instead. We can treat each task as a separate market, and the network operator seeks to maximize its total revenue, and thus profit, across all markets. If the network operator supplies completions of the $i^\text{th}$ task at a price of $x_i$ per unit task completion rate, then users only purchase completion of the $i^\text{th}$ task if they derive a utility of more than $x_i$ per unit task completion rate.

Operating the quantum network is commercially viable if total revenue exceeds total costs. Total revenue depends on the shape of the utility functions $u_i$. Note that, in general, total revenue is lower than the quantum network utility $U_{QN}$. If the cost of providing quantum network services exceeds the total revenue but is lower than the quantum network utility, then operating the quantum network is socially desirable but not commercially viable. In this case, public investment in quantum network services would be essential.

Even if the quantum network is commercially viable, private provision may still not be allocatively efficient. For general utility functions $u_i(p_i)$, neither of the monopolistic markets necessarily achieves the optimal allocation.

Nevertheless, we can show that if the utility functions $u_i(p_i)$ are linear in $p_i$, i.e. if $u_i(p_i) = \alpha_i p_i$ for some $\alpha_i > 0$, then a market in which a profit-maximizing, monopolistic network operator sells task completions achieves the optimal resource allocation $p_i^*$. (This result would apply to the utilities described in Section~\ref{subsec:utilitydef}.) However, a market where entanglement is the quantum network service to be sold may not achieve the optimal resource allocation. We would need $\alpha_i$ to also be proportional to the amount of entanglement consumed per task completion for the entanglement market to achieve the optimal allocation.


\subsection{Measuring utility}

Our framework invites an economic interpretation of the quantum network utility as the social benefit derived from the use of the quantum network. To accurately measure this social benefit, we would need to obtain empirically meaningful measures of the utility users derive from quantum network services.

Theoretically-motivated prescriptions of the utility functions $u_i(p_i)$ may not be realistic in practice. For example, in Section~\ref{sec:qnu_computing}, we used the quantum volume $v = \beta^{\min(\lvert \mathcal{M}_i \rvert, d_i)}$ as a measure of a computation task's utility. An argument could be made that this prescription is not wholly appropriate, as it implies that the utility of a computation task increases exponentially with the number of memories involved, assuming the error rate is sufficiently low. This exponential scaling is unsustainable: the users of a quantum network may be willing to pay twice as much to perform distributed quantum computing over three nodes instead of two, but they may not be willing to pay twice as much to compute over twenty-one nodes instead of twenty. A possible fix is to taper off large quantum volumes above some $v_0$ in the expression for the quantum network utility, using the modified volume $v / (1+v/v_0)$ in place of $v = \beta^{\min(\lvert \mathcal{M}_i \rvert, d_i)}$ in the optimization problem~\eqref{eq:prob}. However, this fix does not resolve fundamental issues with utility functions that have not been empirically validated.

Moreover, the quantum network utility should also take into account other network-enabled applications, such as agreement protocols~\cite{Epping_2017}, distributed sensing~\cite{Khabiboulline_2019} and other known~\cite{Kimble_2008, Wehner_2018} and unknown applications, but there are few convincing theoretical motivations for the utility functions for these tasks.

Therefore, we propose two distinct but non-exclusive principles for measuring utility in practice.

The first principle is to consider the cost of alternatives to using a quantum network. For example, instead of communicating using quantum-network-distributed secret keys, one could employ quantum-safe classical cryptography, or purchase insurance and accept the risk of eavesdropper attacks. The prices of these alternatives are relatively more well-established than the value of quantum key distribution.

The second principle is to estimate the demand curve, which describes the relationship between prices and the quantity of quantum network services demanded. Demand curves allow us to recover the utility derived from quantum network services. Even though there are many practical approaches to demand estimation~\cite{Hausman_1994, Nevo_2001}, it is likely to be difficult in the absence of established markets for quantum network services. However, demand estimation will be more viable as quantum networks become more mature.

\subsection{Market structure}

When quantum network services are provided by private operators, the number of operators and the nature of their competition, i.e. the market structure, determine the resource allocation achieved by the quantum network. Some market structures allocate resources efficiently between competing tasks and users, whereas others could be allocatively inefficient or commercially unviable. The legal framework supporting the private or public provision of quantum network services could also affect market outcomes, as it did in classical telecommunications networks~\cite{Hausman_2013}.

Even though a legal and economic analysis of different market structures is beyond the scope of this paper, we believe that the development of an ecosystem to allocate quantum communication resources efficiently will be beneficial.

\section{Conclusion}

We have presented a general framework for benchmarking quantum networks, starting from the rate region as a description of a network's capabilities. The resulting aggregate utility metric not only takes realistic errors into account, but also facilitates comparison across quantum networks with different nodes and/or technologies. The aggregate utility metric can also be interpreted as the social value provided by the quantum network to its users.

We have also developed an example of an aggregate utility metric for distributed quantum computing. This metric extends the quantum volume from quantum devices to quantum networks. We develop some theoretical and numerical results for the quantum network utility in prototypical quantum networks, and illustrate its scaling behavior with and without gate errors.

The detailed examples we developed account for the value users derive from quantum communication through distributed quantum computing. Our framework is designed to incorporate further applications of quantum communication, such as key distribution and sensing, simply by specifying the utility associated with completing these other tasks. The framework can also be used to guide the design of quantum networks, by choosing the hardware and protocols that maximize the completion of tasks users value most. 

We believe that the adoption of the quantum network utility framework and related aggregate utility metrics will facilitate forecasts of the value of quantum networks at different levels of maturity, and help design quantum networks to maximize their near-term and long-term impacts.

\section*{Acknowledgements}
We thank Gerald Adams (SpaceX), Professor Saikat Guha (University of Arizona), Professor Liang Jiang (University of Chicago) and Dr. Jay M. Gambetta (IBM) for helpful discussions. This work was supported in part by the ARO MURI W911NF2110325, the NSF Center for Quantum Networks, the NSF grant CNS-1955834, and the National Science Foundation to the Computing Research Association for the CIFellows 2020 Program.

\clearpage

\bibliography{references}

\end{document}


\title{Supplementary Information for ``Quantum Network Utility: A Framework for Benchmarking Quantum Networks"}

\author{Yuan Lee}
\thanks{Equal contribution}
\affiliation{Department of Electrical Engineering and Computer Science, Massachusetts Institute of Technology, Cambridge, Massachusetts 02139, USA}

\author{Wenhan Dai}
\thanks{Equal contribution}
\affiliation{Quantum Photonics Laboratory, Massachusetts Institute of Technology, Cambridge, Massachusetts 02139, USA}
\affiliation{College of Information and Computer Sciences, University of Massachusetts, Amherst, MA 01003, USA}

\author{Don Towsley}
\affiliation{College of Information and Computer Sciences, University of Massachusetts, Amherst, MA 01003, USA}

\author{Dirk Englund}
\email{englund@mit.edu}
\affiliation{Department of Electrical Engineering and Computer Science, Massachusetts Institute of Technology, Cambridge, Massachusetts 02139, USA}
\affiliation{Research Laboratory of Electronics, Massachusetts Institute of Technology, Cambridge, Massachusetts 02139, USA}

\date{October 14, 2022}

\maketitle

\appendix

\renewcommand{\appendixname}{Supplementary Note}
\renewcommand{\thesection}{\arabic{section}}
\renewcommand{\thesubsection}{\alph{subsection}}
\counterwithout{equation}{section}

\section{Fundamental Network Description} \label{app:rate_region}

\subsection{The rate region}

As described in the main text, the rate region is a fundamental description of a quantum network's capabilities. The rate region of a quantum network is defined to be the set of quantum communication rates that can be simultaneously enabled by the quantum network. It incorporates information on the quality of its physical links, the efficiency of its entanglement swaps and the throughput of its network routing protocols.

Let the quantum network be described by the graph $(\mathcal{V}, \mathcal{E}_p)$, where $\mathcal{V}$ is the set of nodes and $\mathcal{E}_p$ is the set of physical links. Let the set of communication channels enabled by the quantum network be $\mathcal{E}$. For our purposes, we can think of $\mathcal{E}$ as a set of bipartite or multipartite entangled states, indexed by the nodes over which the state extends. $\mathcal{E}$ describes the output entanglement that can be produced by the quantum network in order to perform tasks. Note that $\mathcal{E}_p \subseteq \mathcal{E}$, because $\mathcal{E}_p$ is the set of entangled states produced by the quantum network without requiring entanglement swaps. In the case where the quantum network only produces bipartite entanglement, $\mathcal{E} = \lbrace \lbrace a, b \rbrace \, \vert \, a, b \in \mathcal{V}, a\neq b \rbrace$ describes the set of all output entangled states. However, in general, communication channels can be between groups of more than two users.

Let $R = (r_s)_{s\in \mathcal{E}}$ be a vector of communication rates, where $r_s$ represents the communication rate through channel $s \in \mathcal{E}$. Let $E = (e_s)_{s \in \mathcal{E}}$ be a vector of communication errors, where $e_s$ represents the average error probability for each unit of communication enabled by channel $s$.

We say a pair $(R, E)$ is feasible for a quantum network if the quantum network can simultaneously sustain communication across every channel $s \in \mathcal{E}$ at a rate $r_s$ with at most error $e_s$. Distillation protocols decrease the error $e_s$ at the cost of lowering the communication rate $r_s$, reflecting a trade-off between the rate and fidelity of a communication channel. When this rate-fidelity tradeoff is not central to the analysis, as is the case in this paper, we assume that the quantum network simply guarantees a maximum error probability $\bar{e}$, so that all error vectors $E = (e_s)_{s\in \mathcal{E}}$ associated with feasible $(R, E)$ have $e_s \leq \bar{e}$ for all $s \in \mathcal{E}$.

The output of a quantum network is described by a vector of communication rates and its associated vector of errors. We define the rate region $\mathcal{Y}$ to be the set of $(R, E)$ that is feasible for the quantum network. When we abstract from the rate-fidelity tradeoff, the projected rate region $\mathcal{Y}\vert_{\bar{e}}$ is relevant, where $\mathcal{Y} \vert_{\bar{e}}$ is defined to be the set of rate vectors $R$ such that $(R, \bar{E}) \in \mathcal{Y}$, and $\bar{E}$ is the error vector with $\bar{e}$ in each component.

\subsection{The feasible task region}

On the other hand, we measure the performance of the quantum network through the tasks that it performs. The feasible task region $\mathcal{W}$ is key to computing the quantum network utility, as it describes the task completion rate vectors that can be attained by the network.

One motivation for using tasks to evaluate the performance of quantum networks is that the nature of a task is independent of the underlying quantum network that completes the task. Different quantum networks can perform the same task, allowing us to compare these quantum networks based on a common benchmark. 

Another motivation is the fact that we derive utility from quantum networks through their applications in computing, communication or sensing. These applications are often associated with well-established markets, and the value of each application is better-understood than the value of the raw entanglement produced by the quantum network. The quantum network utility allows us to attach a value to the raw entanglement produced by the quantum network, which can help a network operator allocate entanglement resources between competing applications. Furthermore, understanding the quantum network utility can also guide the development of efficient markets for quantum network services.

Yet another reason to compute quantum network utilities through tasks is to allow arbitrarily many applications to be incorporated in the same framework. Instead of naively comparing each quantum-enabled application and its classical alternative independently, policymakers can use the quantum network utility to obtain a holistic understanding of the benefits of a quantum network. The quantum network utility can also inform the design of future quantum networks: individual quantum network nodes can be located where they maximize the utility of the quantum network. The application each node should serve can also be decided in a similar way.

As in the main text, let $D$ be the number of tasks the quantum network performs. The quantum network seeks to complete these tasks as frequently as possible. In the general framework, we treat these tasks in an abstract way, but Section~III provides an example of a concrete specification of such tasks. Each task is specific to a coalition of nodes.

Let $P = (p_i)_{i=1}^D$ be a vector of task completion rates, where $p_i$ is the rate at which the $i^{\text{th}}$ task is completed. The feasible task region $\mathcal{W}$ is defined to be the set of task completion rate vectors $P$ that can be sustained by the quantum network. As in the main text, we ssay that a task completion rate vector can be sustained if the rate and error vectors corresponding to the entanglement the quantum network must consume lies within its rate region.

To expand on this definition, let $T(R, E)$ be the set of task completion rate vectors $P$ that can be achieved if the quantum network sustains communication rates $R$ and errors $E$. This function depends on the nature of the task to be completed, and how the tasks are performed using the entanglement provided by the network. Then a vector of task completion rates $P$ can be sustained if there exists $(R, E) \in \mathcal{Y}$ such that $P \in T(R, E)$.

The feasible task region can thus be described as $\mathcal{W} = \bigcup_{(R, E) \in \mathcal{Y}} T(R, E)$. When we abstract from the rate-fidelity tradeoff and fix a maximum error rate $\bar{e}$, the feasible task region is $\mathcal{W} = \bigcup_{R \in \mathcal{Y}_{\bar{e}}} T(R, \bar{E})$.

In summary, the rate region summarizes the fundamental capabilities of a quantum network, whereas the feasible task region describes what the quantum network can do for its users. The correspondence $T(R, E)$ connecting these primitives captures how end users use the entanglement provided by the network.

\newpage
\section{More on the Quantum Network Utility}
\label{app:qnu_universality}

\subsection{The definition}

In our definition of the quantum network utility, we implicitly assume that the value users derive from completing the $i^{\text{th}}$ task does not depend on the rate at which other tasks are completed. Then the utility function $u_i:\mathbb{R}_+ \rightarrow \mathbb{R}$ describes how the utility derived from the repeated completion of the $i^{\text{th}}$ task depends on the rate $p_i$ at which it is completed. The total utility derived from the task completion rate vector $P$ is $\sum_{i=1}^D u_i(p_i)$.

When computing the quantum network utility, we choose the optimal feasible task completion rate vector. This is equivalent to choosing an optimal communication rate vector $R$ in the network's rate region, then choosing to allocate the output entanglement between tasks in a way that maximizes the aggregate utility. In other words, the quantum network utility is
\begin{equation} \label{eqnapp:qnu_definition}
    U_{QN} = \max_{P \in \mathcal{W}} \sum_{i=1}^D u_i(p_i) = \max_{(R, E) \in \mathcal{Y}} \max_{P\in T(R, E)} \sum_{i=1}^D u_i(p_i),
\end{equation}
where $\mathcal{Y}$ and $T(R, E)$ are defined in Supplementary Note~\ref{app:rate_region}.

\subsection{Universality of quantum network utility}

It turns out that the aforementioned assumption (that $u_i$ does not depend on $p_j$ for $j\neq i$) is not restrictive. Under reasonable axioms, any aggregate preferences over task completion rate vectors can be represented by an aggregate utility of the form given in Supp.~Eq.~\eqref{eqnapp:qnu_definition}.

Aggregate preferences are described by the binary relation $\succsim$. If the group of all quantum network users is better off when tasks are completed according to rate vector $P$ rather than $P'$, then we say that the users prefer $P$ to $P'$ on aggregate, i.e. $P\succsim P'$. Such preference relations are often used as a primitive in economic welfare analysis. A preference relation characterizes how users as a whole value different combinations of task completion rates.

We introduce the following definitions from the economics and decision theory literature.
\begin{definition}[Utility representation]
We say that a utility function $u:\mathbb{R}_+^D \rightarrow \mathbb{R}$ represents a preference relation $\succsim$ if the following statement holds:
\begin{equation*}
    u(P) \geq u(P') \Leftrightarrow P \succsim P' \quad \forall P, P' \in \mathbb{R}_+^D.
\end{equation*}
\end{definition}

\begin{remark}
In economics, the utility function is an ordinal quantity: it only describes an ordering across task completion rate vectors. Any positive monotone transformation applied to the utility function preserves the preference relation it describes. However, as we will discuss later, we can treat the utility function as a cardinal quantity in special cases.
\end{remark}

\begin{definition}[Additive utility representation]
We say that a preference relation $\succsim$ admits an additive utility representation if it can be represented by a utility function $u:\mathbb{R}_+^D \rightarrow \mathbb{R}$ of the form
\begin{equation*}
    u(P) = u(p_1, \ldots, p_D) = \sum_{i=1}^D u_i(p_i)
\end{equation*}
where $u_i:\mathbb{R}_+ \rightarrow \mathbb{R}$ ($i=1, \ldots, D$) are single-variable functions.
\end{definition}

\begin{remark}
In an additive utility representation, the contribution of the $i^\text{th}$ task to the aggregate utility does not depend on the rate at which other tasks are completed. The aggregate utility of the form given in Supp.~Eq.~\eqref{eqnapp:qnu_definition} is an additive utility.
\end{remark}

We now introduce five axioms for aggregate preferences. These axioms will be described in terms of task completion rate vectors, but they also apply to other bundles of goods whose quantities are components of a ``rate vector".
\begin{enumerate}
    \item \textbf{Completeness.} For any pair of task completion rate vectors $P, P' \in \mathbb{R}_+^{D}$, either $P\succsim P'$ or $P'\succsim P$ holds.
    \item \textbf{Transitivity.} If $P \succsim P'$ and $P' \succsim P''$, then $P \succsim P''$.
    \item \textbf{Continuity.} If $\lbrace P^{(n)} \rbrace_{n\in \mathbb{N}}$ and $\lbrace Q^{(n)} \rbrace_{n\in \mathbb{N}}$ are two convergent sequences in $\mathbb{R}_+^D$ with $P^{(n)} \succsim Q^{(n)}$ for all $n \in \mathbb{N}$, then $\lim_{n\rightarrow \infty} P^{(n)} \succsim \lim_{n\rightarrow \infty} Q^{(n)}$ as well. Here, the limits are defined based on Euclidean distance.
    \item \textbf{Positiveness.} $P\succsim 0$ for all $P \in \mathbb{R}_+^D$. Here, $0$ denotes the zero rate vector in $\mathbb{R}_+^D$, i.e. when the network does not complete any task.
    \item \textbf{Preferential independence.} For any subset $A \subseteq \lbrace 1, \ldots, D \rbrace$, let $P_A = (p_i)_{i \in A}$ be the restriction to $A$ of the task completion rate vector $P = (p_i)_{i=1}^D$. Also let $\bar{A}$ be the complement of $A$. Then, for any subset of tasks $A \subseteq \lbrace 1, \ldots, D \rbrace$ and any task completion rate vectors $P, P' \in \mathbb{R}_+^D$, $(P_A, P_{\bar{A}}) \succsim (P'_A, P_{\bar{A}})$ if and only if $(P_A, P'_{\bar{A}}) \succsim (P'_A, P'_{\bar{A}})$.
\end{enumerate}

The first four axioms are uncontroversial for any reasonable preferences. Axioms 1 (completeness) and 2 (transitivity) are basic requirements for users to act rationally based on their knowledge of their utilities. Axiom 3 (continuity) means that the users' aggregate preferences are not subject to wild fluctuations. Axiom 4 (positiveness) states that the completion of tasks provides value, and the non-completion of tasks provides no value.

Axiom 5 (preferential independence) means that preferences among any subset of tasks do not depend on the rates at which the complementary subset of tasks are performed. This axiom is less obviously true, and in fact Axiom 5 may not hold because of complementary tasks: the rate at which some task is delivered may affect users' preferences for other tasks. For example, consider a network that can perform three tasks $\lbrace t_1, t_2, t_3 \rbrace$. We may prefer $t_2$ to $t_3$ if $t_1$ is unavailable. However, if $t_1$ and $t_3$ are complementary (because they can together be used for some other valuable application), then we may prefer $t_1 + t_3$ to $t_1 + t_2$. This would violate the preferential independence assumption.

Nonetheless, Axiom 5 is a reasonable assumption if the set of tasks the network can perform is sufficiently rich. For example, extending the situation provided above, the complementary uses of $t_1$ and $t_3$ can be captured by a larger task $t_4$ shared between all users who derive value from the complementary tasks. An application in which $t_1$ and $t_3$ are jointly used is equivalent to an application using $t_4$. If $t_4$ is in the set of tasks, then preferences over $t_1$ and $t_3$ should only account for applications in which the tasks are enjoyed separately, so as to avoid double-counting the value derived from applications where $t_4$ is used. Including $t_4$ in the set of tasks would restore preferential independence.

The following theorem states that, under axioms 1 to 5, any aggregate preferences over task completion rate vectors can be represented by an aggregate utility $\sum_{i=1}^D u_i(p_i)$ for some utility functions $\lbrace u_i \rbrace_{i=1}^D$.

\begin{theorem}
Assume the number of tasks $D$ is finite. The aggregate preferences $\succsim$ admit an additive utility representation if Axioms 1, 2, 3 and 5 hold. Moreover, if at least three tasks affect preferences, then Axioms 1, 2, 3 and 5 are also necessary conditions for the existence of an additive utility representation.
\end{theorem}
\begin{proof}
This result follows from Debreu's additive utility theorem~\cite{Debreu_1983}.
\end{proof}

Therefore, the implicit assumption we make in writing the aggregate utility as $\sum_{i=1}^D u_i(p_i)$ holds if the set of $D$ tasks the quantum network can perform is a comprehensive one.

Now, we make the claim that the quantum network utility is a universal measure of the value provided by a quantum network.

\begin{definition}
We say that a preference relation $\succsim$ has a universal collection of utility representations if, for any pair of utility functions $u, v:\mathbb{R}_+^D\rightarrow \mathbb{R}$ in the collection, there exists a positive real number $\alpha > 0$ such that $u = \alpha v$.
\end{definition}
\begin{remark}
Universality means that the preference relation has a utility representation that is unique up to a scaling factor. This scaling factor can be interpreted as a unit conversion, like the factor of $10^3$ when converting between kilometers and meters. Therefore, the utility representation can be treated as a universal measure of value when attached to appropriate units.
\end{remark}

We have the following result, also from Ref.~\cite{Debreu_1983}.

\begin{theorem}
Assume the number of tasks $D$ is finite and Axioms 1, 2, 3 and 5 hold, so that the aggregate preferences $\succsim$ admit an additive utility representation. Then $u(P) = \sum_{i=1}^D u_i(p_i)$ and $v(P) = \sum_{i=1}^D v_i(p_i)$ are both additive utility representations of $\succsim$ if and only if there exist some constants $\alpha > 0$ and $\beta$ such that $u_i(p) = \alpha v_i(p) + \beta$ for all $p \in \mathbb{R}_+$ and $i \in \lbrace 1, \ldots, D\rbrace$.
\end{theorem}

In other words, the utility functions $\lbrace u_i \rbrace_{i=1}^D$ are unique up to a positive affine transformation. If we further assume Axiom 4, then $u_i(p) \geq u_i(0)$ for all $p > 0$ and $i \in \lbrace 1, \ldots, D \rbrace$. Note that $u_i$ is not necessarily an increasing function over its domain.

It is natural to set the utility of the zero rate vector to be zero, indicating that no value is derived from the network when no tasks are complicated. Then $u_i(0) = 0$ for all $i \in \lbrace 1, \ldots, D \rbrace$. Moreover, under this restriction, the additive utility representation is unique up to a scaling factor.

Hence, for any preferences satisfying Axioms 1 to 5, the collection of additive utility representations with zero utility at zero rate is universal. The aggregate utility derived from a task completion rate vector can be used as a universal measure of its value.

Consequently, we can treat the quantum network utility as a universal measure of a quantum network's performance: if measured in the same units, the magnitude of the quantum network utility can be used to directly compare different networks.

\newpage
\section{More on the Feasible Task Region for Distributed Quantum Computing} \label{app:HOG_task}

\subsection{Derivation}

We explain how we arrive at the feasible task region $\mathcal{W}$ for distributed quantum computing, as described in the optimization problem~(3) in the main text.
We do so in two steps. First, we map the rate region $\mathcal{Y}\vert_{\bar{e}}$ to the feasible task region $\mathcal{W}$ through the correspondence $T(R, \bar{E})$. ($\mathcal{Y}\vert_{\bar{e}}$ and $T(R, \bar{E})$ are defined in Supplementary Note~\ref{app:rate_region}.) Then, we construct the rate region $\mathcal{Y} \vert_{\bar{e}}$ itself.

For the first step, recall that one layer in the HOG task involves applying random SU(4) gates to pairs of memories that are matched at random. On average, every pair of memories has the same probability of being matched. An $m$-memory, depth-$d$ HOG task involves $d\lfloor m/2 \rfloor$ SU(4) gates. Therefore, completing such a HOG task requires every pair of memories in the coalition to be matched $2d \lfloor m/2 \rfloor / m(m-1)$ times in expectation.

We can implement an arbitrary SU(4) gate over two distant nodes using at most two Bell states connecting those nodes. In particular, we can teleport one of the involved qubits from its node to the other, perform the required SU(4) gate locally, and teleport the corresponding qubit back to its original node.\footnote{Even if ancilla memories are not available, at most a constant number of Bell states are needed to implement an arbitrary SU(4) gate. In particular, any SU(4) gate can be implemented with at most three CNOT gates~\cite{Vatan_2004}, on top of other single-qubit gates, and each CNOT gate can be teleported using one Bell state~\cite{Jiang_2007}.}

This means that completing a depth $d_i$ HOG task over coalition $\mathcal{M}_i \subseteq \mathcal{V}$ at rate $p_i$ consumes entanglement from every pair of nodes in the coalition at rate
\begin{equation*}
    r^i = p_i \times 2 \times \frac{2d_i \lfloor \lvert \mathcal{M}_i \rvert/2 \rfloor}{\lvert \mathcal{M}_i \rvert (\lvert \mathcal{M}_i \rvert-1)} = \begin{cases}
    2 p_i d_i \left(\lvert \mathcal{M}_i \rvert - 1\right)^{-1} & \text{if } \lvert \mathcal{M}_i \rvert \text{ is even,} \\
    2 p_i d_i \lvert \mathcal{M}_i\rvert^{-1} & \text{if } \lvert \mathcal{M}_i \rvert \text{ is odd.}
    \end{cases}
\end{equation*}
To achieve the task completion rate $P = (p_i)_{i=1}^{D}$, the network will consume entanglement $(a, b) \in \mathcal{E}$ at rate
\begin{equation*}
    r_{ab} = \sum_{i=1}^{D} r^i \mathbbm{1}_{a \in \mathcal{M}_i} \mathbbm{1}_{b \in \mathcal{M}_i},
\end{equation*}
where the indicator function $\mathbbm{1}$ is defined as 
\begin{equation*}
\mathbbm{1}_X = \begin{cases} 1 & \text{if } X \text{ is true}, \\ 0 & \text{otherwise.} \end{cases}
\end{equation*}

In this first step, it remains to determine if the HOG task $(\mathcal{M}_i, d_i)$ can be performed with sufficiently high fidelity. Applying the approximation in Eq.~(2) from the main text, we require $\lvert\mathcal{M}_i\rvert d_i \leq 1/\epsilon_{\text{eff}}$ as a constraint for any task that is completed with positive rate. As two Bell states are required to perform an arbitrary SU(4) two-qubit gate and Bell states are produced with at most error $\bar{e}$ by assumption, the effective error probability is $\epsilon_{\text{eff}} = 2\bar{e}$ to first order.

Hence, task completion rates $P = (p_i)_{i=1}^{D}$ lie in $T(R, \bar{E})$ for rate vector $R = (r_{ab})_{(a, b) \in \mathcal{E}}$ if
\begin{gather*}
    r_{ab} \geq \sum_{i=1}^{D} 2 p_i d_i \left\lbrace \begin{aligned}
    \left(\lvert \mathcal{M}_i \rvert - 1\right)^{-1} & \text{if } \lvert\mathcal{M}_i\rvert \text{ is even} \\
    \lvert \mathcal{M}_i \rvert^{-1} & \text{if } \lvert \mathcal{M}_i \rvert \text{ is odd}
    \end{aligned}
    \right\rbrace \mathbbm{1}_{a \in \mathcal{M}_i} \mathbbm{1}_{b \in \mathcal{M}_i} \, \forall \, a, b \in \mathcal{V}, \\ 
    \lvert \mathcal{M}_i \rvert d_i \leq \frac{1}{\epsilon_{\text{eff}}} \, \forall \, i = 1, \ldots, D.
\end{gather*}

For the second step, we follow Ref.~\cite{Dai_2020} and assume that entanglement swapping occurs with a given efficiency $q_c$ in node $c \in \mathcal{V}$. (This includes the probability that entanglement swapping is successful, and also any operations needed to keep the error of output entanglement below $\bar{e}_0$.) We assume that, in the absence of any entanglement swaps, the network produces entanglement $(a, b) \in \mathcal{E}$ at rate $f_{ab}$. Note that if nodes $a$ and $b$ are not connected by a physical link, then any communication between these nodes must be generated using entanglement swaps, so $f_{ab} = 0$.

Then, as in Ref.~\cite{Dai_2020}, the rate vector $R = (r_{ab})_{(a, b) \in \mathcal{E}}$ is in the rate region $\mathcal{Y}\vert_{\bar{e}}$ if there exist variables $w_{ab}^{ac}$ for all $a, b, c \in \mathcal{V}$ such that
\begin{gather*}
    r_{ab} \leq f_{ab} + \sum_{c \in \mathcal{V} \setminus \lbrace a, b \rbrace} q_c \left( \frac{w_{ab}^{ac} + w_{ab}^{bc}}{2} \right) - \sum_{c \in \mathcal{V} \setminus \lbrace a, b \rbrace} \left( w_{ac}^{ab} + w_{bc}^{ab} \right) \, \forall \, a, b \in \mathcal{V}, \\
    w_{ab}^{ac} = w_{ab}^{bc} \geq 0 \, \forall \, a, b, c \in \mathcal{V}.
\end{gather*}

Intuitively, the variables $w_{ab}^{ac}$ represent the rate at which $(a, c)$ entanglements are used to generate $(a, b)$ entanglements via entanglement swaps at node $c$. The inequality above imposes an ``entanglement conservation" condition for each node pair, accounting for entanglement from physical links and for the efficiency of entanglement swaps. This rate region is valid when each node has sufficiently many perfect memories allocated for communication. Otherwise, this rate region is an approximation of the true rate region of the network.

Therefore, the feasible task region $\mathcal{W}$ is the set of task completion rate vectors $P = (p_i)_{i=1}^D$ where there exist $R = (r_{ab})_{(a, b) \in \mathcal{E}}$ and $(w_{ab}^{ac})_{a\neq b\neq c \in \mathcal{V}}$ such that the following holds: 
\begin{align*}
    & \, p_i \geq 0 \, \forall \, i, \quad r_{ab} \geq 0 \, \forall \, a, b \in \mathcal{V}, \quad w_{ab}^{ac} \geq 0 \, \forall \, a, b, c \in \mathcal{V}; \\
    & \, r_{ab} = \sum_{i=1}^{D} 2 p_i d_i \left\lbrace \begin{aligned}
    \left(\lvert \mathcal{M}_i \rvert - 1\right)^{-1} & \text{if } \lvert\mathcal{M}_i\rvert \text{ is even} \\
    \lvert \mathcal{M}_i \rvert^{-1} & \text{if } \lvert \mathcal{M}_i \rvert \text{ is odd}
    \end{aligned}
    \right\rbrace \mathbbm{1}_{a \in \mathcal{M}_i} \mathbbm{1}_{b \in \mathcal{M}_i} \, \forall \, a, b \in \mathcal{V}; \\
    & \, \lvert \mathcal{M}_i \rvert d_i \leq \frac{1}{\epsilon_{\text{eff}}} \, \forall \, i = 1, \ldots, D \text{ such that } p_i > 0; \\
    & \, r_{ab} \leq f_{ab} + \sum_{c \in \mathcal{V} \setminus \lbrace a, b \rbrace} q_c \left( \frac{w_{ab}^{ac} + w_{ab}^{bc}}{2} \right) - \sum_{c \in \mathcal{V} \setminus \lbrace a, b \rbrace} \left( w_{ac}^{ab} + w_{bc}^{ab} \right) \, \forall \, a, b \in \mathcal{V}; \\
    & \, w_{ab}^{ac} = w_{ab}^{bc} \, \forall \, a, b, c \in \mathcal{V}.
\end{align*}

\subsection{Technical simplification} \label{app_sub:HOG_depth}

So far, we have allowed the set of $D$ tasks to include HOG tasks ($i$ and $j$) over the same coalition of nodes ($\mathcal{M}_i = \mathcal{M}_j$) but with different depths ($d_i \neq d_j$). It is straightforward to see that all tasks involving the same coalition of nodes should be performed to the same depth, which is the depth that has the largest potential contribution to the utility. (This observation follows from the fact that HOG tasks over the same coalition consume entanglement in the same relative ratios across states, regardless of depth.) Therefore, the set of tasks only has to include HOG tasks over distinct coalitions, such that each coalition $\mathcal{M}_i$ is associated with an optimal depth $d_i$. Restricting the set of HOG tasks to distinct coalitions of nodes reduces $D$, and thus the size of the linear program, significantly.

With the change of variables $\tilde{p}_i = p_i d_i$, we can show that the optimal depth $d_i$ for a coalition $\mathcal{M}_i$ is given by
\begin{equation*}
    d_i = \argmax_{1 \leq d \leq d_{\text{max}}} \frac{\beta^d}{d},
\end{equation*}
where $d_{\text{max}} = \min(\lvert\mathcal{M}_i \rvert, 1/\sqrt{\epsilon_{\text{eff}}} \lvert\mathcal{M}_i \rvert)$.

\newpage
\section{Propositions for Repeater Chains} \label{app:prop_proofs}

In this section, we provide proofs of the propositions stated in Section~III~D. We also restate the propositions below.

\begin{proposition}\label{propapp:connected_coalition}
In the optimization problem~(3) (from the main text), there exists an optimal solution such that for any task with $p_i > 0$, the corresponding coalition $\mathcal{M}_i$ is connected.
\end{proposition}

Therefore, when finding the task completion rate vector that maximizes the aggregate utility derived from the repeater chain, we only have to give tasks associated with connected coalitions a non-zero task completion rate.

We use the following lemma in the proof of Proposition~\ref{propapp:connected_coalition}. For the ease of exposition, we introduce the binary relation $\succeq$ between rate vectors. We say that $R \succeq R'$ if, for any possible quantum network, $R$ is in the rate region only if $R'$ is also in the capacity region.

\begin{figure}[ht]
    \centering
    \includegraphics[width = 0.8\textwidth]{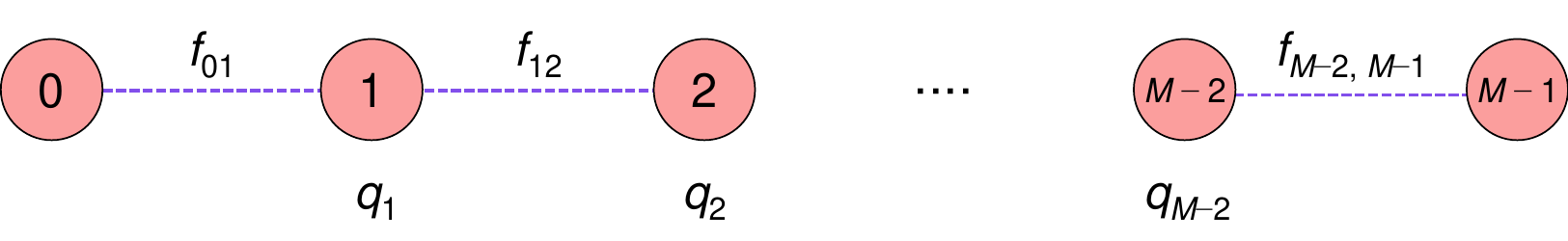}
    \caption{Illustration of a length-$M$ quantum repeater chain. The nodes are $\mathcal{V} = \lbrace 0, 1, \ldots, M-1 \rbrace$. The no-swap communication rates are $f_{ab}$ for $a, b \in \mathcal{V}$, and the efficiency of entanglement swapping at node $c \in \mathcal{V}$ is $q_c$.}
    \label{figapp:repeater_chain}
\end{figure}

\begin{lemma} \label{lemmaapp:succeq}
Suppose $R = (r_{ab})$ is in the rate region of the quantum repeater chain shown in Supp.~Fig.~\ref{figapp:repeater_chain}. Let arbitrary $\ell, k, j \in \mathcal{V}$ and $r \geq 0$ be such that $ \ell < k < j$ and $r \leq r_{\ell j}$. Define the rate vector $R' = (r'_{ab})$ by
\begin{equation*}
    r_{ab}' = \begin{cases}
    r_{ab} - r & \text{if } \lbrace a, b \rbrace = \lbrace \ell, j \rbrace, \\
    r_{ab} + r & \text{if } \lbrace a, b \rbrace = \lbrace \ell, k \rbrace \text{ or } \lbrace a, b \rbrace = \lbrace k, j \rbrace, \\
    r_{ab} & \text{otherwise}.
    \end{cases}
\end{equation*}
Then $R \succeq R'$.
\end{lemma}

\begin{figure}[ht]
    \centering
    \includegraphics[width = 0.8\textwidth]{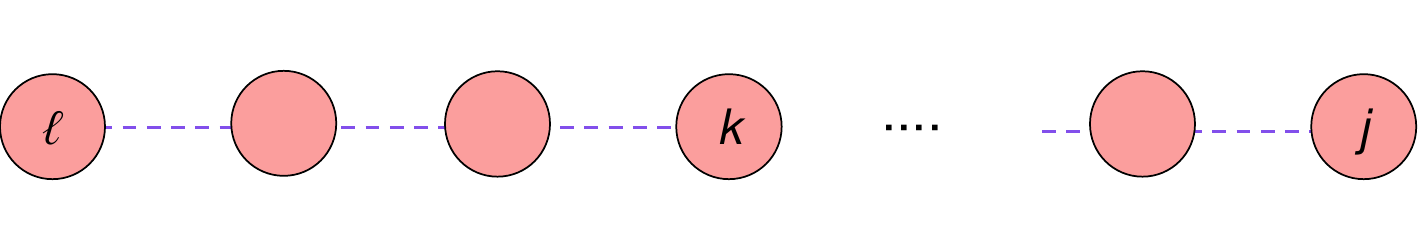}
    \caption{Relative locations of $\ell$, $k$ and $j$ in Lemma~\ref{lemmaapp:succeq}.}
    \label{figapp:lkj}
\end{figure}

\begin{proof}
The relative locations of $\ell$, $k$ and $j$ are shown in Supp.~Fig.~\ref{figapp:lkj}. For any protocol that achieves the rate vector $R$, we make the following modification of the protocol: whenever node $k$ needs to perform entanglement swapping to distribute entanglement between $\ell$ and $j$ (either directly or as an intermediate step) according to the original protocol, node $k$ instead does nothing with probability $r / r_{\ell j}$ in the new protocol. In this way, if in the original protocol an entangled pair between $\ell$ and $j$ is generated, then in the modified protocol, two entangled pairs (one $(\ell, k)$ and another $(k, j)$) are generated with probability $r/r_{\ell j}$, and an entangled pair $(\ell, j)$ is generated with probability $1 - r/r_{\ell j}$. Therefore, in the new protocol, the entanglement rate between nodes $\ell$ and $j$ becomes $r'_{\ell j} = r_{\ell j} (1- r / r_{\ell j}) = r_{\ell j} - r$. The entanglement rate between nodes $\ell$ and $k$ becomes $r'_{\ell k} = r_{\ell k} + r$, and that between nodes $k$ and $j$ becomes $r_{kj} + r$.
\end{proof}

\begin{proof}[Proof of Proposition~\ref{propapp:connected_coalition}]

Let $a$ be the leftmost node in a coalition $\mathcal{M}_i$. The node to the right of $a$ is $a+1$, the node to the right of $a+1$ is $a+2$, and so on. Let $b$ be the rightmost node in the largest connected subset of $\mathcal{M}_i$ that contains $a$. The remaining $K$ nodes in $\mathcal{M}_i$ are denoted as $c_1, c_2, \ldots, c_K$. By definition, $c_1 > b+1$.

Without loss of generality, we assume $b-a+1 \leq K$. Otherwise, we consider the rightmost node in $\mathcal{M}_i$ and its corresponding largest connected subset instead.

Now consider the alternative coalition $\mathcal{M}_h$ comprising nodes $a+1, a+2, \ldots, b, b+1, c_1, c_2, \ldots c_K$. Let $R = (r_{ab})$ be the entanglement rate required to support the task completion rate $P = (p_j)_{j=1}^D$. Let $R' = (r'_{ab})$ be the entanglement rate required to support the task completion rate $P' = (p'_j)_{j=1}^D$, where
\begin{equation*}
    p'_j = \begin{cases}
        0 & \text{if } j = i. \\
        p_h + p_i & \text{if } j = h, \\
        p_j & \text{otherwise}.
    \end{cases}
\end{equation*}
In other words, the network now performs tasks over coalition $\mathcal{M}_h$ instead of $\mathcal{M}_i$, while fixing the completion rate for other tasks.

We now show that $R \succeq R'$, so if a network can supply the entanglement needed to complete the $i^\text{th}$ task, it can instead supply the entanglement needed for the $h^\text{th}$ task.

To perform the $h^\text{th}$ task instead of the $i^\text{th}$ task, we need less entanglement between node pairs in $\mathcal{A} = \lbrace (a, a+1), (a, a+2), \ldots, (a, b), (a, c_1), (a, c_2), \ldots, (a, c_K) \rbrace$ and more entanglement between node pairs in $ \mathcal{A}' = \lbrace (b+1, a+1), (b+1, a+2), \ldots (b+1, b), (b+1, c_1), (b+1, c_2), \ldots, (b+1, c_K) \rbrace$. The absolute change in the required entanglement rate for any of the node pairs listed above is
\begin{equation*}
    r = 2 p_i d_i  \left\lbrace \begin{aligned}
    \left(\lvert \mathcal{M}_i \rvert - 1\right)^{-1} & \text{if } \lvert\mathcal{M}_i\rvert \text{ is even} \\
    \lvert \mathcal{M}_i \rvert^{-1} & \text{if } \lvert \mathcal{M}_i \rvert \text{ is odd}
    \end{aligned}
    \right\rbrace,
\end{equation*}
where $d_i$ is the HOG task depth for the $i^\text{th}$ task, as defined in Supplementary Note~\ref{app_sub:HOG_depth}. Note that the HOG task depth for the $h^\text{th}$ task is also $d_i$ because $\lvert \mathcal{M}_i \rvert = \lvert \mathcal{M}_h \rvert$. 

Then consider an intermediate rate vector $R'' = (r''_{ab})$, where $\mathcal{A}'' = \lbrace (a+1, c_1), (a+2, c_2), \ldots, (b, c_{b-a}), (b+1, c_{b-a+1}), (b+1, c_{b-a+2}), \ldots (b+1, c_K) \rbrace$ and
\begin{equation*}
    r''_{ab} = \begin{cases}
    r'_{ab} - r & \text{if } (a, b) \in \mathcal{A}', \\
    r'_{ab} + r & \text{if } (a, b) \in \mathcal{A}'', \\
    r'_{ab} & \text{otherwise}.
    \end{cases}
\end{equation*}
Iteratively applying Lemma~\ref{lemmaapp:succeq} with $\ell = a+s$, $k = b+1$ and $j \in \lbrace 1, 2, \ldots, b-a \rbrace$ implies that $R'' \succeq R'$. It is straightforward to verify that $R \succeq R''$. This shows that $R \succeq R'$, i.e. the task over coalition $\mathcal{M}_h$ requires less entanglement than that over $\mathcal{M}_i$ when achieving the same task completion rate.

Note that the index of the leftmost node in $\mathcal{M}_h$ is strictly greater than in $\mathcal{M}_i$. If $\mathcal{M}_h$ is not connected, we can keep finding a new coalition in the same way until a connected coalition is found. At most $c_K - a + 1 -M$ steps would be required before finding a connected coalition. This connected coalition requires less entanglement than $\mathcal{M}_i$ while having the same task completion rate. As this connected coalition has the same number of nodes as $\mathcal{M}_i$, it has the same contribution to the quantum network utility. In particular, $\sum_{j=1}^D u_j(p_j) = \sum_{j=1}^D u_j(p_j')$ for the utility functions $u_j$ defined in Section~III~B based on the quantum volume.

The optimization problem~(3) (from the main text) has at least one optimal solution. If there is a positive task completion rate over a non-connected coalition, following the procedure above gives an optimal solution where only connected coalitions have positive task completion rates. This completes the proof.
\end{proof}

For a quantum repeater chain with $M$ nodes, the number of connected coalitions is $M(M-1)/2$. This is much smaller than the number of subsets of $\mathcal{V}$, which grows exponentially with $M$. Therefore, Proposition~\ref{propapp:connected_coalition} allows us to efficiently solve the optimization problem~(3).

Proposition~\ref{propapp:connected_coalition} also offers a ``byproduct" result that provides a lower bound on the size of the largest coalition.

For simplicity, we consider homogeneous repeater chains, which is a special case of quantum repeater chains in which $f_{b, b+1} = f$ for all $b \in \lbrace 0, 1, \ldots, M-2 \rbrace$ and $q_c = q$ for all $c \in \lbrace 1, 2, \ldots, M-2 \rbrace$.

\begin{proposition}\label{propapp:size_largest_coalition}
In a homogeneous repeater chain with perfect quantum memories and no gate errors (i.e. $\epsiloneff = 0$), the size of the largest coalition with nonzero task rate in an optimal solution is bounded from below by
\begin{align*}
M + \log_\beta \frac{M^{\log q} }{(1+q)M^3(M-1)^2/4}.
\end{align*}
\end{proposition}

\begin{proof}
Consider the optimal solution. (If there are multiple optimal solutions, consider the one that has the largest coalition.) Let $M^*$ denote the size of the largest coalition with a positive task rate. Following the proof of Proposition \ref{propapp:connected_coalition}, we can assume the optimal solution corresponds to connected coalitions. Evidently, the optimal value is upper bounded by
\begin{align}\label{eqapp:upper_bound}
    f \beta^{M^*} M(M-1)/2.
\end{align}

We now consider a sub-optimal solution that uses only one coalition, $\mathcal{V}$. We divide each physical link between adjacent nodes into $M(M-1)/2$ sub-channels, such that with probability $2/M(M-1)$, this link is used to generate entanglement between nodes $a$ and $b$, $\forall a, b $ with $0\leq a < b \leq M-1$. Using Claim 1 in Ref.~\cite{Dai_2020}, we can show that the pairwise entanglement rate can be lower bounded by
\begin{align*}
 \frac{ f  M^{\log q} }{(1+q)M(M-1)/2}.
\end{align*}
Therefore, the HOG task involving all nodes in $\mathcal{V}$ has a rate at least
\begin{align*}
 \frac{ f  M^{\log q} }{(1+q)M^2(M-1)/2}
\end{align*}
and the corresponding utility is 
\begin{align*}
 \frac{ f \beta^M M^{\log q} }{(1+q)M^2(M-1)/2}.
\end{align*}
This value is upper bounded by Supp.~Eq.~\eqref{eqapp:upper_bound}, giving the desired result.
\end{proof}

Proposition \ref{propapp:size_largest_coalition} states that in a homogeneous repeater chain used for distributed quantum computing, the size of the largest coalition increases as $M-O(\log{M})$. Thus, for large $M$, if the memories and quantum gates are perfect, almost all the nodes in the network should be in the same coalition. 

We next consider the case of a network that produces entanglement with imperfect fidelity. The size of the largest coalition no longer increases as $M-O(\log{M})$.

\begin{proposition}\label{propapp:lb_size_lc_error}
In a quantum network with errors (i.e. $\epsiloneff > 0$), the size of the largest coalition with nonzero task rate in an optimal solution is bounded from above by $\lfloor 1/\sqrt{\epsiloneff}\rfloor$.
\end{proposition}

\begin{proof}
Suppose the size of the largest coalition in an optimal solution is greater than $\lfloor 1/\sqrt{\epsiloneff}\rfloor$. Let $m^*$ denote this size. Let $d^*$ and $p^*$ denote the HOG task depth and the task completion rate corresponding to the largest coalition respectively. If $d^*\geq m^*$, then 
\begin{align*}
     m^* d^* \geq (\lfloor 1/\sqrt{\epsiloneff}\rfloor +1)^2> 1/\epsiloneff
\end{align*}
which contradicts Eq.~(2) (from the main text). Therefore $d^* \leq m^*-1$.

We then consider a $d^*$-node subset of the largest coalition. Let $\mathcal{M}_0$ denote such a node set and let $\mathcal{M}_1$ denote the node set consisting of the remaining nodes in the largest coalition. By definition, $|\mathcal{M}_0| =d^*$ and $|\mathcal{M}_1|>0$. 

We now consider a new solution based on the supposed optimal solution. We replace the task associated with the largest coalition with a task over coalition $\mathcal{M}_0$ at depth $d^*$ and task completion rate $p^*$. We also perform a new two-node task at positive rate: one node is in $\mathcal{M}_0$ and the other is in $\mathcal{M}_1$. Note that this new solution is feasible because less entanglement is used relative to the supposed optimal solution. However, the utility of the new solution is strictly greater than that of the original solution, because the utility of $\mathcal{M}_0$ is the same as that of the original largest coalition and the utility of the new two-node coalition is positive. This contracdicts the optimality of the orginal solution, showing that $m^*\leq \lfloor 1/\sqrt{\epsiloneff}\rfloor$.
\end{proof}

\begin{proposition}\label{propapp:ub_size_lc_error}
In a homogeneous repeater chain with errors, the size of the largest coalition with nonzero task rate in an optimal solution is bounded from below by
\begin{align*}
m+\log_\beta{\frac{4m^{\log_2{q}}\lfloor M/m\rfloor}{(1+q)m^3(m-1)(2M-m+1)}}
\end{align*}
where $m=\lfloor\sqrt{1/\epsiloneff}\rfloor$.
\end{proposition}
\begin{proof}
Consider the optimal solution. (If there are multiple optimal solutions, consider the one that has the largest coalition.) Let $M^*$ denote the size of the largest coalition. Following Proposition~\ref{propapp:connected_coalition}, we assume the optimal solution corresponds to connected coalitions. By Proposition~\ref{propapp:lb_size_lc_error}, the number of coalitions that could possibly have positive task rate is $(2M-m+1)m/2$. The optimal aggregate utility is then upper bounded by
\begin{align}\label{eqapp:upper_bound_2}
    f \beta^{M^*} m(2M+m-1)/2.
\end{align}

We now consider a sub-optimal solution that uses $\lfloor M/m\rfloor$ coalitions, denoted by $\mathcal{M}_i$, $i=0,1,2, \dotsc, \lfloor M/m\rfloor-1$. These coalitions are non-overlapping node sets. Specifically, coalition $\mathcal{M}_i$ comprises nodes $im +j$, $j=0,1,2,\dotsc, m-1$.

Now consider coalition $\mathcal{M}_i$. We divide each physical link between nodes $j$ and $j+1$ in this coalition into $m(m-1)/2$ sub-channels, so that the link is used to generate entanglement between nodes $j$ and $k$ ($im \leq j \leq kim+m-1$) with probability $2/m(m-1)$. Using Claim 1 in Ref.~\cite{Dai_2020}, we can show that the pairwise entanglement rate can be lower bounded by
\begin{align*}
 \frac{ f  m^{\log q} }{(1+q)m(m-1)/2}.
\end{align*}
Therefore, the computing task involving all the nodes in $\mathcal{M}_i$ has a rate at least
\begin{align*}
 \frac{ f  m^{\log q} }{(1+q)m^2(m-1)/2}
\end{align*}
and the corresponding aggregate utility involving all the coalitions is 
\begin{align*}
 \lfloor M/m\rfloor\frac{ f \beta^m m^{\log q} }{(1+q)m^2(m-1)/2}.
\end{align*}
This value is  upper bounded by Supp.~Eq.~\eqref{eqapp:upper_bound_2}, giving the desired result.
\end{proof}

Proposition~\ref{propapp:lb_size_lc_error} states that the largest coalition can have at most $m = \lfloor 1/\sqrt{\epsilon_{\text{eff}}} \rfloor$ nodes. Proposition~\ref{propapp:ub_size_lc_error} states that the size of the largest coalition is asymptotically close to this upper bound. In particular, for sufficiently large $M$ and sufficiently small $\epsiloneff$, the size of the largest coalition increases as $m - O(\log m)$.

\newpage

\begin{table}[ht!]
    \centering
    \caption{Summary of terminology and notation.}
    \label{tab:notation}
    \begin{tabular}{r l}
        \textbf{Name} & \textbf{Explanation} \\ \hline \hline
        Communication channels & Bi-/multi-partite entanglement shared between users \\ \hline
        Groups of users & Users who are interested in / derive value from \vspace{-6pt} \\
        & some application of quantum communication \\ \hline
        Coalitions of nodes & Nodes over which a task is performed \\ \hline
        $\mathcal{E}$ & Set of channels \\ \hline
        $\mathcal{E}_p$ & Set of physical channels \\ \hline
        $D$ & Total number of tasks \\ \hline
        Rate vector $R$ & Entanglement rates $r_s$ for each channel \\ \hline
        Error vector $E$ & Errors $e_s$ for each channel \\ \hline
        Task completion rates $P$ & Task rates $p_i$ for each task \\ \hline
        Rate region $\mathcal{Y}$ & Set of feasible rate and error vectors $(R, E)$ \\ \hline
        Rate region $\mathcal{Y}\vert_{\bar{e}}$ & Set of feasible rate vectors $R$ with maximum error $\bar{e}$ \\ \hline
        Feasible task region $\mathcal{W}$ & Set of feasible task completion rates \\ \hline
        $T(R, E)$ & Feasible task region given rate and error vectors $R, E$ \\ \hline
        $T(R, \bar{E})$ & Feasible task region given rate $R$ and maximum error $\bar{e}$ \\ \hline
        $\bar{E}$ & Error vector with error $\bar{e}$ for each channel \\ \hline
        $u_i(p_i)$ & Utility function for the $i^\text{th}$ task, given task rate $p_i$ \\ \hline
        $U_{QN}$ & Quantum network utility \\ \hline
        $v$ & Quantum volume \\ \hline
        $U_{\text{comp}}$ & Quantum network utility for distributed quantum computing \\ \hline
        $\beta$ & Base of quantum volume \\ \hline
        $\mathcal{V}$ & Set of network nodes \\ \hline
        $\mathcal{M}_i$ & Coalition of nodes associated with the $i^\text{th}$ task, subset of $\mathcal{V}$ \\ \hline
        $d_i$ & Depth of HOG computation for the $i^\text{th}$ task \\ \hline
        $\epsilon_{\text{eff}}$ & Effective error per two-qubit gate \\ \hline
        $r^i$ & Entanglement rate consumed by the $i^\text{th}$ task \\ \hline
        $r_{ab}$ & Entanglement rate between nodes $a, b \in \mathcal{V}$ \vspace{-6pt} \\ 
        & (equivalent to $r_s$, where $s = (a, b)$) \\ \hline
        $f_{ab}$ & Entanglement rate generated by physical links between \vspace{-6pt} \\
        & $a, b \in \mathcal{V}$, also known as the no-swap rate \\ \hline
        $w_{ab}^{ac}$ & Entanglement flows, from Ref.~\cite{Dai_2020} \\ \hline
        $M$ & Number of nodes in a network = $\lvert \mathcal{V} \rvert$ \\ \hline
        Chain length & Number of nodes in a repeater chain \\ \hline
        Spoke size $M_{\text{side}}$ & Number of spokes on each side of the bar \vspace{-6pt} \\
        & in a dumbbell network (so $M = 2M_{\text{side}} + 2$) \\ \hline
        Entanglement graph & Graphical representation of a rate vector $R$ \\ \hline
    \end{tabular}
\end{table}

\clearpage
\bibliography{references}